\documentclass[11pt, a4paper]{article}
\usepackage[utf8]{inputenc}
\usepackage{graphicx}
\usepackage{amsmath}
\usepackage{mathtools}
\usepackage{amssymb}
\usepackage{fullpage}
\usepackage{array, booktabs}
\usepackage{comment}
\usepackage{jinstpub}
\usepackage[sort&compress,numbers]{natbib}
\usepackage{lineno}
\usepackage{dcolumn}
\usepackage{appendix}
\usepackage{siunitx}
\usepackage{url}
\usepackage{caption}
\usepackage{subcaption}
\usepackage{multirow}

\begin{document}
\title{Calibration Plan for the SBC 10-kg Liquid Argon Detector with 100 eV Target Threshold}

% The SBC Collaboration Author List

\note[*]{Corresponding author.}

%alphabetical affiation list?
\affiliation[a]{Department of Physics, University of Alberta, Edmonton, T6G 2E1, Canada}
\affiliation[b]{Department of Physics and Astronomy, University of California Riverside, Riverside, California 92507, USA}
\affiliation[c]{Department of Physics, University of California Santa Barbara, Santa Barbara, California, 93106, USA}
\affiliation[d]{Department of Physics, Drexel University, Philadelphia, Pennsylvania 19104, USA}
\affiliation[e]{Fermi National Accelerator Laboratory, Batavia, Illinois 60510, USA}
\affiliation[f]{Department of Physics, Indiana University South Bend, South Bend, Indiana 46634, USA}
%\affiliation{Department of Chemical Engineering and Materials Science, University of Minnesota, Minneapolis 55455, USA}
\affiliation[g]{D\'{e}partement de Physique, Universit\'{e} de Montr\'{e}al, Montr\'{e}al, H3T 1J4, Canada}
\affiliation[h]{Instituto de F\'{\i}sica, Universidad Nacional Aut\'onoma de M\'exico, A.P. 20-364, Ciudad de M\'exico 01000, M\'exico}
\affiliation[i]{Northeastern Illinois University, Chicago, Illinois 60625, USA}
\affiliation[j]{Department of Physics and Astronomy, Northwestern University, Evanston, Illinois 60208, USA}
\affiliation[k]{Department of Physics, Queen's University, Kingston, K7L 3N6, Canada}
\affiliation[l]{SNOLAB, Lively, Ontario, P3Y 1N2, Canada}
\affiliation[m]{TRIUMF,  Vancouver,  BC  V6T  2A3,  Canada}

\author[h]{E.~Alfonso-Pita}
%\affiliation{Instituto de F\'{\i}sica, Universidad Nacional Aut\'onoma de M\'exico, A.P. 20-364, Ciudad de M\'exico 01000, M\'exico}

\author[e]{D.~Baxter}
%\affiliation{Fermi National Accelerator Laboratory, Batavia, Illinois 60510, USA}

\author[f]{E.~Behnke}
%\affiliation{Department of Physics, Indiana University South Bend, South Bend, Indiana 46634, USA}

%\author[d]{M.~Bressler}
%\altaffiliation[Present address: ]{University of Massachusetts, Amherst, 01002, USA.}
%\affiliation{Department of Physics, Drexel University, Philadelphia, Pennsylvania 19104, USA}

\author[k]{B.~Broerman}
%\affiliation{Department of Physics, Queen's University, Kingston, K7L 3N6, Canada}

\author[k]{K.~Clark}
%\affiliation{Department of Physics, Queen's University, Kingston, K7L 3N6, Canada}

\author[k]{J.~Corbett}
%\affiliation{Department of Physics, Queen's University, Kingston, K7L 3N6, Canada}

\author[e]{M.~Crisler}
%\affiliation{Fermi National Accelerator Laboratory, Batavia, Illinois 60510, USA}
%\affiliation{Pacific Northwest National Laboratory,\ Richland,\ Washington\ 99354,\ USA}

\author[j,e]{C.~E.~Dahl}
%\affiliation{Department of Physics and Astronomy, Northwestern University, Evanston, Illinois 60208, USA}
%\affiliation{Fermi National Accelerator Laboratory, Batavia, Illinois 60510, USA}

\author[k]{K.~Dering}
%\affiliation{Department of Physics, Queen's University, Kingston, K7L 3N6, Canada}

\author[k]{A.~de St.~Croix}
%\affiliation{Department of Physics, Queen's University, Kingston, K7L 3N6, Canada}

\author[a]{D.~Durnford}
%\affiliation{Department of Physics, University of Alberta, Edmonton, T6G 2E1, Canada}

\author[m]{P.~Giampa}
%\affiliation{TRIUMF,  Vancouver,  BC  V6T  2A3,  Canada}

\author[l,1]{J.~Hall\nolinebreak
\note{Now at The Fedoruk Centre.}}
%\affiliation{SNOLAB, Lively, Ontario, P3Y 1N2, Canada}

\author[i]{O.~Harris}
%\affiliation{Northeastern Illinois University, Chicago, Illinois 60625, USA}

\author[k]{H.~Hawley-Herrera}
%\affiliation{Department of Physics, Queen's University, Kingston, K7L 3N6, Canada}

\author[c,2]{L.~Joseph\nolinebreak
\note{Now at University of Washington.}}
%\affiliation{Department of Physics, University of California Santa Barbara, Santa Barbara, California, 93106, USA}

\author[j]{A.~Kucich}
%\affiliation{Department of Physics and Astronomy, Northwestern University, Evanston, Illinois 60208, USA}

\author[d,*]{N.~Lamb}
%\note[*]{Corresponding author.}}
\emailAdd{nrl47@drexel.edu}
%\affiliation{Department of Physics, Drexel University, Philadelphia, Pennsylvania 19104, USA}

\author[g]{M.~Laurin}
%\affiliation{D\'{e}partement de Physique, Universit\'{e} de Montr\'{e}al, Montr\'{e}al, H3T 1J4, Canada}

\author[f]{I.~Levine}
%\affiliation{Department of Physics, Indiana University South Bend, South Bend, Indiana 46634, USA}

\author[c]{W.~H.~Lippincott}
%\affiliation{Department of Physics, University of California Santa Barbara, Santa Barbara, California, 93106, USA}

\author[j]{B.~Mitra}
%\affiliation{Department of Physics and Astronomy, Northwestern University, Evanston, Illinois 60208, USA}

\author[d]{R.~Neilson}
%\affiliation{Department of Physics, Drexel University, Philadelphia, Pennsylvania 19104, USA}

\author[j]{O.~Nicholson}
%\affiliation{Department of Physics and Astronomy, Northwestern University, Evanston, Illinois 60208, USA}

\author[a]{M.-C.~Piro}
%\affiliation{Department of Physics, University of Alberta, Edmonton, T6G 2E1, Canada}

\author[e]{G.~Putnam}
%\affiliation{Fermi National Accelerator Laboratory, Batavia, Illinois 60510, USA}

\author[d]{D.~Pyda}
%\affiliation{Department of Physics, Drexel University, Philadelphia, Pennsylvania 19104, USA}

\author[j]{Z.~Sheng}
%\affiliation{Department of Physics and Astronomy, Northwestern University, Evanston, Illinois 60208, USA}

\author[k]{G.~Sweeney}
%\affiliation{Department of Physics, Queen's University, Kingston, K7L 3N6, Canada}

\author[h]{O.~Vald\'{e}s-Martínez}

\author[h]{E.~V\'{a}zquez-J\'{a}uregui}
%\affiliation{Instituto de F\'{\i}sica, Universidad Nacional Aut\'onoma de M\'exico, A.P. 20-364, Ciudad de M\'exico 01000, M\'exico}

%\author{D.~Velasco}
%\affiliation{Department of Physics and Astronomy, Northwestern University, Evanston, Illinois 60208, USA}

\author[b]{S.~Westerdale}
%\affiliation{Department of Physics and Astronomy, University of California Riverside, Riverside, California 92507, USA}

\author[c]{T.~J.~Whitis}
%\affiliation{Department of Physics, University of California Santa Barbara, Santa Barbara, California, 93106, USA}

\author[d,3]{S.~Windle\nolinebreak
\note{Now at University of Maryland.}}

\author[k]{A.~Wright}
%\affiliation{Department of Physics, Queen's University, Kingston, K7L 3N6, Canada}

\author[k]{E.~Wyman}
%\affiliation{Department of Physics, Queen's University, Kingston, K7L 3N6, Canada}

\author[c,*]{R.~Zhang}
\emailAdd{runzezhang@ucsb.edu}
%\affiliation{Department of Physics, University of California Santa Barbara, Santa Barbara, California, 93106, USA}

%\author[h]{A.~Zu\~niga-Reyes}
% \altaffiliation[Present address: ]{University of Toronto, Toronto, M5S 1A7, Canada.}
%\affiliation{Instituto de F\'{\i}sica, Universidad Nacional Aut\'onoma de M\'exico, A.P. 20-364, Ciudad de M\'exico 01000, M\'exico}

\collaboration{SBC collaboration}

	%\noaffiliation
%	\email{sbc@snolab.ca}

%\author[1]{N. Lamb,R.Zhang}
%\emailAdd{nrl47@drexel.edu,runzezhang@ucsb.edu}
%\note[1]{Corresponding Author}
%\author{R. Neilson}
%\author{E. Alphonso-Pita}
%%\author{The Drexel PICO group}
%\affiliation{Department of Physics, Drexel University\\ Philadelphia, Pennsylvania, USA}
\date{\today}

\abstract{The Scintillating Bubble Chamber (SBC) Collaboration is designing a new generation of low background, noble liquid bubble chamber experiments with sub-keV nuclear recoil threshold. These experiments combine the electronic recoil blindness of a bubble chamber with the energy resolution of noble liquid scintillation, and maintain electron recoil discrimination at higher degrees of superheat (lower nuclear recoil thresholds) than Freon-based bubble chambers. A 10-kg liquid argon bubble chamber has the potential to set world leading limits on the dark matter nucleon cross-section for $\mathcal{O}$(GeV/$c^{2}$) masses, and to perform a high statistics coherent elastic neutrino nuclear scattering measurement with reactor neutrinos. This work presents a detailed calibration plan to measure the detector response of these experiments, combining photoneutron scattering with two new techniques to induce sub-keV nuclear recoils: nuclear Thomson scattering and thermal neutron capture.
}

\maketitle
\flushbottom
\section{Introduction}
The Scintillating Bubble Chamber (SBC) Collaboration is developing noble liquid bubble chambers as low energy, low background nuclear recoil detectors. A scalable detector technology sensitive to 0.1 - 10 keV nuclear recoils (NR) that is blind to electronic recoils (ER) would have substantive physics reach for dark matter searches and neutrino physics~\cite{SBCuniverse,SBCsnowmass,SBCcevns}. A prototype noble liquid bubble chamber has demonstrated these characteristics, with scintillation light providing an additional detection channel that can primarily be used as a veto for high-energy backgrounds~\cite{SBCxenon1st}.

Bubble chambers use superheated liquid to detect particle interactions \cite{Seitz,glaser}. If an interaction creates a local hot spot exceeding both the energy and energy density thresholds set by the degree of superheat in the target, a microscopic bubble is formed.  This bubble rapidly grows, generating an acoustic signal and enabling macroscopic bubble photography. In the past, bubble chambers have been operated as tracking chambers for beam experiments~\cite{Gargamelle,BEBC}. More recently, bubble chambers have found use in low background rare event searches, such as dark matter experiments~\cite{Picasso1stResults,COUPP1stresult,60complete}. In this mode, bubble chambers generally operate as threshold detectors (either detecting an interaction or not), although MeV-scale alpha decays can be discriminated from keV-scale NRs acoustically~\cite{PICASSOalpha}. Crucially, energy deposits from ERs are typically too diffuse to nucleate bubbles, leading to ER blindness. Freon bubble chambers with NR detection thresholds of $\sim$3~keV and ER sensitivities of $\mathcal{O}(10^{-9})$ have set world leading constraints on spin dependent dark matter-proton couplings for dark matter masses above 3~GeV/$c^{2}$~\cite{60complete}.

The SBC Collaboration is constructing two 10 kg liquid argon bubble chambers with a target NR threshold of 100 eV, more than an order of magnitude below the threshold of Freon bubble chambers or any other existing liquid-based ER-discriminating detector. These chambers include a dark matter search experiment at SNOLAB (SBC-SNOLAB) and a dedicated calibration detector at Fermilab (SBC-LAr10). Detection of sub-keV argon NRs would provide sensitivity to $\mathcal{O}$(GeV/$c^{2}$) mass dark matter, a mass range favored by lower mass WIMPs and asymmetric dark matter models \cite{ParticleDM,DarkMatter,SnowmassFrontier}. The strongest existing limits on 1 GeV/$c^{2}$ mass dark matter nucleon cross-sections are set a bit below $10^{-40}$ cm$^{2}$ by the DarkSide collaboration \cite{DarkSide-50:2023fcw}. With sub-keV thresholds, low backgrounds, and a 10 kg-year exposure, SBC could explore cross-sections of approximately 10$^{-43}$ cm$^2$ \cite{SBCuniverse,SBCsnowmass}. A future ton-scale experiment could constrain the dark matter-nucleon cross-section down to the solar neutrino fog \cite{neutrino_fog}.

A 10 kg SBC detector also has the potential to observe Coherent Elastic Neutrino Nuclear Scattering (CE$\nu$NS) from reactor neutrinos. The CE$\nu$NS interaction has a high cross section and can provide insights into neutrino and beyond standard model physics \cite{CEvNS,CEvNSimp}. The first CE$\nu$NS measurement was made by the COHERENT collaboration in 2017 using a CsI detector at the Spallation Neutron Source (SNS) \cite{FirstCoherent}. The lower-energy (sub-keV) NRs from reactor CE$\nu$NS evaded detection until a 3.7 sigma observation by the CONUS collaboration in 2025, using high-purity germanium detectors at a nuclear power plant in Liebstadt, Switzerland \cite{CONUS_25}. Reactors provide a high flux of neutrinos, and the reactor flux is pure anti-$\nu_{e}$, in contrast to the mixed flavor flux from stopped pion sources such as the SNS. With a 100 eV NR threshold, an argon bubble chamber could observe $\sim$1 CE$\nu$NS events/kg/day at 3~m from a 1 MW$_{\text{th}}$ reactor or $\sim$15 events/kg/day at 30~m from a 2000 MW$_{\text{th}}$ reactor \cite{SBCcevns}. Due to the ER blindness of liquid noble bubble chambers, SBC could achieve a signal to background ratio of roughly 10 in these scenarios.

A calibration of the NR energy threshold for bubble nucleation is necessary to reach SBC physics objectives and is the subject of this paper. The aim is to reach a 20\% uncertainty on the NR threshold to support the dark matter goals of SBC \cite{SBCsnowmass}. Constraining the weak mixing angle and the neutrino magnetic moment through a reactor CE$\nu$NS measurement requires a threshold uncertainty of $\sim$5\% \cite{SBCcevns}. In order to meet these calibration goals, new techniques are presented, including nuclear Thomson scattering by gamma rays and thermal neutron capture, along with established techniques using elastic neutron scattering from photoneutron sources~\cite{YBeCollar,BePhotoneutron,pico_calibration,SBCxenonNRER}. The potential for low-threshold dark matter detectors to observe nuclear Thomson scattering was noted in Ref.~\cite{AlanCoherentPhotons}. Nuclear Thomson scattering can be induced by $\mathcal{O}$(MeV) gamma rays, with the low energy NRs produced being ideal for calibrating at detection thresholds below 300 eV.  Thermal neutron capture can create low energy NRs when the absorbing nucleus relaxes by gamma ray emission \cite{neutron_capture_cal_theory,CRAB}. Neutron capture on argon results in $\mathcal{O}$(100~eV) NRs, accompanied by scintillation light produced by gamma ray interactions that can be used to tag the event as a neutron capture, resulting in a high-purity sample of calibration events.

The calibration strategy described below calls for a combined plan: Thomson scattering and photoneutron scattering are best suited for NR thresholds below 300 eV; neutron capture and photoneutrons are best suited for thresholds of $\gtrsim$ 300 eV. The following sections will introduce the history of bubble chambers for rare event searches, explain the SBC-LAr10 calibration strategy, and showcase simulated calibrations for SBC liquid argon detectors. 

\section{The Scintillating Bubble Chamber Experiment}

Bubble chambers use superheated liquid to detect particle interactions via bubble formation. The detection threshold is a function of temperature and pressure, and can be approximated by the Seitz “hot-spike” model \cite{Seitz,ERpaper}.  In the Seitz model, proto-bubbles form by heat energy deposited locally, vaporizing the liquid. Above a critical radius, of the order of 10s of nm, gas pressure in the proto-bubble overcomes the surrounding liquid pressure and surface tension, such that the bubble grows to macroscopic size. The total energy required to create a critically sized bubble establishes the thermodynamic energy threshold. Below this threshold, any proto-bubbles formed collapse before they can be observed. Higher temperature and lower pressure increase the superheat, decreasing the threshold for creating a critically-sized bubble.

Operating low background chambers requires high purity active detector fluids, fused silica vessels free of surface contamination, sufficient shielding from external backgrounds, and efficient rejection of ER backgrounds as developed by PICASSO, COUPP and PICO \cite{PICASSOalpha,COUPPNIM,ERpaper,PICO2L,DBCpaper}. Because sensitivity to ER backgrounds is highly dependent on thermodynamic conditions, the lowest detection threshold a rare event bubble chamber can practically achieve is limited by its sensitivity to ER background events. The detection threshold is set by minimizing the ER background rate with respect to the NR acceptance. Below 3 keV thresholds in Freon, bubble rates from ER backgrounds rise steeply. For this reason, PICO's Freon bubble chambers operate with detection thresholds of $\sim$3 keV and above.

%SBC plans to use PICO inspired bubble chambers with noble liquid targets for GeV dark matter and neutrino experiments \cite{SBCuniverse}.
Noble liquid bubble chambers have demonstrated improved ER rejection compared to Freon chambers; a prototype SBC xenon detector demonstrated sensitivity to $\sim$1~keV NRs but saw no evidence of ER nucleation at thermodynamic energy thresholds as low as 0.5~keV~\cite{SBCxenon1st,SBCxenonNR,SBCxenonNRER}. The proposed explanation for the improved ER rejection is that the scintillation mechanism and other forms of radiation in noble liquids carry energy away from the recoil site. In molecular fluids, by contrast, excitation energy is effectively converted to heat locally through molecular degrees of freedom. Based on the success of the xenon prototype, SBC anticipates that a liquid argon chamber will remain blind to ERs at sub-keV thresholds. With spontaneous bubble formation from thermal fluctuations predicted to be negligible in argon at or above a 40~eV threshold~\cite{HomogeneousNucleation}, SBC-LAr10 has a goal of demonstrating sensitivity to 100~eV NRs~\cite{SBCuniverse,SBCsnowmass}.

\begin{figure*}
    \centering
    \includegraphics[width=\textwidth]{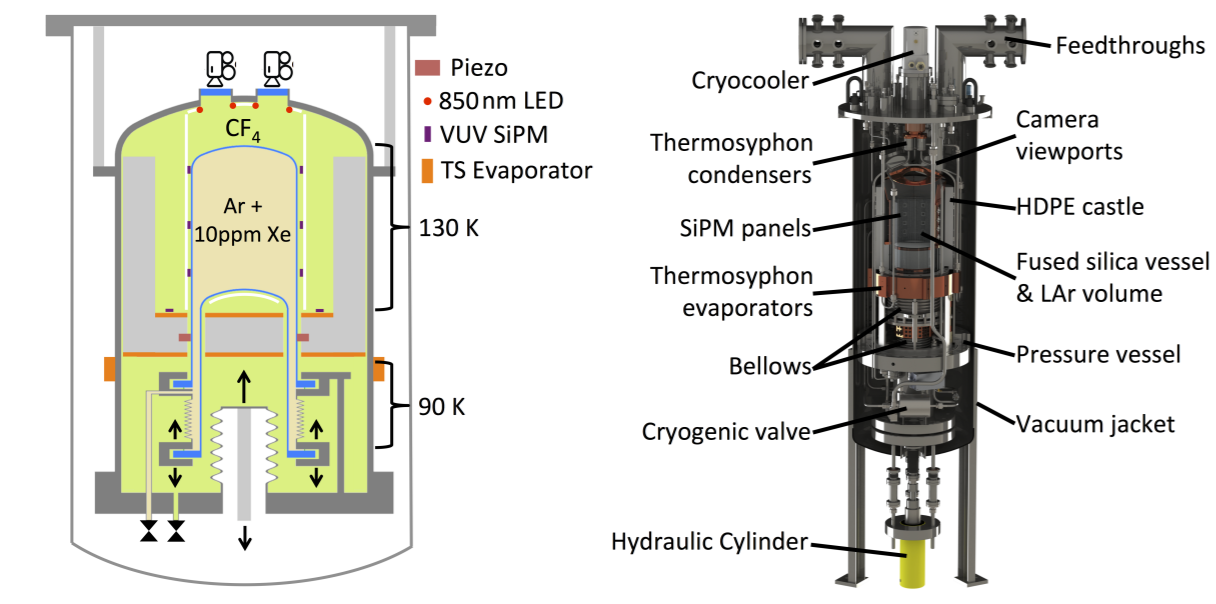}
    \caption{(Left) The SBC-LAr10 design shown as a schematic diagram, showing the warmer superheated liquid (argon doped with 10~ppm xenon) kept at 130~K between the fused silica jars and the colder 90~K stable liquid region. The schematic also highlights the cameras and piezo acoustic sensors for bubble detection in addition to the SiPMs for scintillation light collection. (Right) The labeled CAD model for the detector \cite{SBCsnowmass}.}
    \label{fig:bubble-chamber}
\end{figure*}

The design for the SBC-LAr10 bubble chamber is shown in Figure \ref{fig:bubble-chamber}. The liquid argon active detector volume is kept between two fused silica jars, while the inside of the lower jar and the outside of the upper jar are filled with cryogenic carbon tetrafluoride (CF$_{4}$) hydraulic fluid. A stainless steel bellows between the silica vessels maintains pressure balance between the argon and the CF$_{4}$, and a stainless steel pressure vessel contains the liquid volumes. The fluids are cooled by three closed-loop nitrogen thermosyphons in thermal contact with a cryocooler, and the detector is separated into two temperature regions by an insulating high-density polyethylene (HDPE) “castle.” The upper region is superheated at 130 K and is the active detector. The fluid in the lower region is maintained at 90~K, a temperature at which argon is not superheated at the target operating pressure. The pressure vessel is enclosed in a stainless steel vacuum jacket. A recessed stainless steel tube with 4.5~cm diameter extends 29~cm down from the top of the vacuum jacket, allowing for calibration source placement 12.1 cm from the top of the active volume, shown later in Figure~\ref{fig:SourceGeometry}.

Bubbles are detected with cameras, acoustic sensors, and pressure transducers, similar to PICO. Three cameras allow for position reconstruction of bubble location, with LED illumination, and eight piezo acoustic sensors listen for the sound of bubble formation. Scintillation is detected via an array of 32 silicon photomultipliers (SiPMs). The argon is doped with 10 ppm xenon to wavelength shift the scintillation light so that it can pass through the silica vessels. Scintillation is used to veto high energy background events and enable the thermal neutron capture calibration; low energy NRs, as expected from dark matter scattering, are below the scintillation detection threshold. Scintillation detection is only possible when the chamber is dark, so the LEDs will either be operated with a $\sim$10\% duty cycle, or will be triggered only once a bubble is detected.

The chamber is operated through expansion-compression cycles, during which a hydraulic cylinder expands and compresses the active detector volume with a piston. With the chamber initially in the compressed state at a pressure of $\sim$25 bara, the pressure is slowly lowered to $\sim$1.4 bara at constant temperature to superheat the argon in the upper region. The chamber is then ready to detect particle interactions. Once a bubble is detected by the cameras, acoustic sensors, or pressure transducers, the chamber is recompressed to collapse the bubble, restarting the cycle. The time it takes to reset following a bubble is around 30 seconds, setting the maximum detection rate to $\mathcal{O}$(1) bubble event per minute ($\mathcal{O}$(1,000) bubbles per day).

SBC-LAr10 will operate in the MINOS tunnel at Fermilab, with 300 meters of water equivalent overburden to reduce cosmogenic backgrounds. Based on simulations of rock neutrons, rock gamma rays, and muon-induced neutrons, a total fast-neutron background rate of $\sim$2 bubble nucleation events per hour is expected, low enough that no shielding is required for calibration.

\section{Bubble Chamber Calibration}
The Seitz model used to approximate bubble chamber detection thresholds does not fully predict the details of bubble nucleation efficiency of NRs, as it does not account for alternative methods of energy loss, uncertainties in model parameters, and event-by-event variations in the time and distance over which energy is deposited \cite{PICASSOalpha,COUPPCF3I,pico_calibration}. The calculated thermodynamic threshold is used only as an estimate of the NR detection threshold, with a dedicated calibration effort required for each target fluid for more precise understanding of nucleation efficiency.

The probability of a NR forming a bubble as a function of energy deposited is referred to as the nucleation efficiency function. To experimentally measure this function, NRs are induced by radioactive sources or particle beams. By comparing the NR spectrum predicted by simulations with the observed bubble rate, the efficiency function is inferred. Radioactive sources used to measure NR response include photoneutron sources such as yttrium-beryllium ($^{88}$Y-Be) and $\alpha$-neutron sources such as americium-beryllium ($^{241}$Am-Be) \cite{COUPPNIM,PICASSOalpha}. Particle beam calibrations used in the past include the filtered $\sim$100 keV neutron beam from the Portuguese Research Reactor used by SIMPLE \cite{SIMPLEbeam} and the 12 GeV pion beam used by COUPP at Fermilab \cite{COUPPCF3I}. 

A comprehensive calibration of a bubble chamber for $>$ 2 keV NRs has been demonstrated by the PICO collaboration for C$_3$F$_8$ bubble chambers using an SbBe photoneutron source, an $^{241}$Am-Be $\alpha$-neutron source, and a mono-energetic neutron beam at the University of Montr\'eal \cite{pico_calibration}. 
The results suggest that effective nucleation by NRs in C$_{3}$F$_{8}$ requires energy depositions roughly twice the thermodynamic threshold.
%$^{124}$Sb produces gamma rays that can eject nearly mono-energetic 24~keV neutrons from the surrounding $^9$Be. Alphas from the $^{241}$Am source eject neutrons of order 1 MeV. The Montr\'eal accelerator produces beams of 50, 61, and 97 keV neutrons.
%The PICO analysis fits calibration data to simulations with a piecewise efficiency function \cite{pico_calibration}. 
A similar technique using $^{124}$Sb-Be and $^{88}$Y-Be photoneutron sources, along with a $^{252}$Cf fission neutron source, was used for the calibration of the SBC 30-g xenon bubble chamber, demonstrating that sub-keV NR thresholds can be achieved with a noble liquid target while remaining insensitive to ERs and confirming that the effective threshold is still well approximated by the calculated thermodynamic threshold at sub-keV thresholds~\cite{SBCxenonNRER}.

%A similar technique was used for the SBC 30-g xenon bubble chamber calibration, which used SbBe and YBe as sources of 24 keV and 145 keV photoneutron, respectively \cite{SBCxenonNRER}.

The SBC detectors (SBC-LAr10 and SBC-SNOLAB) will operate at the lowest threshold for which the detector is blind to ERs and spontaneous nucleation. The thermodynamic threshold for spontaneous nucleation can be conservatively calculated to be 40 eV, but the threshold for ER blindness is unknown --- to be determined in an initial calibration phase with high-activity gamma ray sources. The NR calibration plan for SBC-LAr10 considers thresholds between 80 and 400 eV.

The calibration strategy involves a comparison of the observed bubble rate from calibration sources to the rate from detailed simulations of the detector geometry. A complete calibration will be performed at one thermodynamic threshold (i.e., fixed pressure and temperature), unlike previous bubble chamber calibrations that involved taking data at several thresholds and making assumptions about the way the nucleation efficiency scales with the thermodynamic threshold~\cite{pico_calibration}. The calibration can be repeated at different thresholds as necessary, measuring the efficiency function at each threshold independently. 

\section{Source Simulations} \label{Source Simulations}

The argon NR spectrum induced by each of three calibration source types --- photoneutron scattering, photon-nucleus elastic scattering (Thomson scattering), and thermal neutron capture --- is simulated in the SBC geometry with Geant4~\cite{geant4}.

\subsection{Photoneutron Scattering} \label{Photoneutron Scattering}
Photoneutron sources consist of radioactive isotopes with decay gamma rays above the neutron binding energy of the target nuclide, typically beryllium ($Q_{\text{Be}}$=1.665 MeV), ejecting neutrons:
\begin{equation}
^{9} \text{Be} + \gamma \xrightarrow{}   \prescript{8}{}{\text{Be}} + n,
\end{equation}
\begin{equation}
E_{\gamma} - Q_{\text{Be}}= T_{n} + T_{\text{Be}},
\end{equation}
where $E_{\gamma}$ is the energy of the inbound gamma, $T_{n}$ is the final kinetic energy of the photoneutron, and $T_{\text{Be}}$ the beryllium kinetic energy. As a two-body final-state reaction, photoneutrons are mono-energetic in the center-of-mass frame, with a small energy spread in the lab frame. The photoneutron sources considered for SBC-LAr10 (listed in Table \ref{table:simrates_pn}) are $^{58}$Co-Be,  $^{124}$Sb-Be, and $^{207}$Bi-Be, producing 9, 24 (rarely 380), and 94~keV neutrons respectively, leading to NRs from neutron scattering in the tens of eV to keV ranges. $^{58}$Co and $^{124}$Sb both have half-lives of $\sim$2 months, but there are no longer-half-life alternatives that produce such low-energy neutrons.

An important consideration is that neutrons can scatter more than once in the detector, losing energy with each scatter, allowing the calibration to fit not only the rate of bubble nucleation but also the number of bubbles in each event. The ratio between single bubbles and higher multiplicity events can provide information on the efficiency function that is independent of source strength.
\begin{table}
\caption{Information on the photoneutron sources considered for use in the calibration. The photoneutron sources produce approximately mono-energetic neutrons, with an energy spread of $\mathcal{O}$(10\%). The max NR energy describes the max recoil energy on argon.}
\begin{tabular}{l l l l l l}
\toprule
%Source Configuration & \multicolumn{2}{c}{ K-Shell Rate} & \multicolumn{2}{c}{L-Shell Rate} & \multicolumn{2}{c}{Energy Deposition Rate} \\
% & \multicolumn{2}{c}{[h$^{-1}$~ppm$^{-1}$]} & \multicolumn{2}{c}{[h$^{-1}$~ppm$^{-1}$]} & \multicolumn{2}{c}{[MeV~s$^{-1}$]} \\
Source & \multicolumn{1}{l}{Primary} & \multicolumn{1}{l}{Branching} & \multicolumn{1}{l}{Mean Neutron} & \multicolumn{1}{l}{Max NR} & \multicolumn{1}{l}{Half-Life}\\
 & \multicolumn{1}{l}{Gammas [MeV]} & \multicolumn{1}{l}{Ratio [\%]} & \multicolumn{1}{l}{Energy [keV]} & \multicolumn{1}{l}{Energy [keV]} & \multicolumn{1}{l}{[years]}\\
\hline
\hline
$^{58}$Co-Be &1.675 &0.52&9.1 &0.94&0.19 \\
$^{124}$Sb-Be &1.691,\ 2.091 &47,\ 5 &24,\ 380 &  2.4,\  37& 0.16\\
$^{207}$Bi-Be &1.770 & 7 & 94 &9.2&31.6 \\
\hline

\bottomrule
\end{tabular}
\label{table:simrates_pn}
\end{table}
\begin{figure}[!ht]
    \centering
    \includegraphics[width=0.6\textwidth]{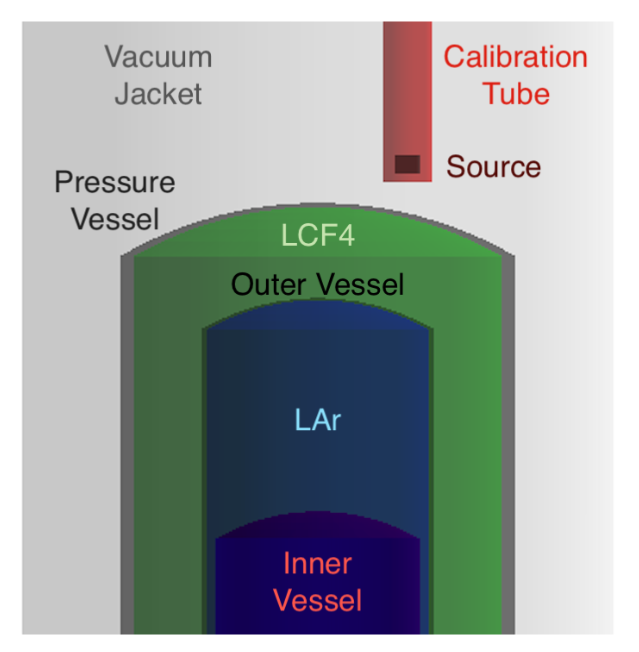}
    \caption{The simulated geometry in the vicinity of the calibration tube and the active detector volume. The bottom of the calibration tube is 12.1 cm from the top of the active LAr. Particles from a calibration source must traverse the 0.6~cm stainless steel cap on the source tube, the 0.95 cm thick stainless steel pressure vessel wall, at least 10 cm of liquid CF$_{4}$, and the 0.51~cm thick fused silica outer vessel to reach the active detector fluid. The simulated geometry also features major detector components, such as the HDPE castle, that are not included in this figure but can be seen in Figure \ref{fig:bubble-chamber}.}
    \label{fig:SourceGeometry}
\end{figure}

Photoneutron sources are simulated in the SBC-LAr10 calibration tube in the Geant4 simulated geometry shown in Figure~\ref{fig:SourceGeometry}. The geometry includes the stainless steel vacuum jacket surrounding a layer of insulating vacuum and the other large components shown in Figure~\ref{fig:bubble-chamber}. The closest point in the active liquid argon is 12.1 cm from the source location. The photoneutron simulations have two phases due to the low cross-section for photoneutron production. The first phase, using SOURCES-4c \cite{Sources4c}, simulates decays in the radioactive sources and neutron generation in the $^{9}$Be. The $^{9}$Be cylinder has a radius 2.38 cm, a height of 7.62 cm, a total mass of 250 g, and the gamma source is placed in the center. The second phase simulates neutron propagation and NRs in the target fluid using Geant4 version 10.3.1.

\begin{figure*}
    \centering
    \includegraphics[width=0.7\textwidth]{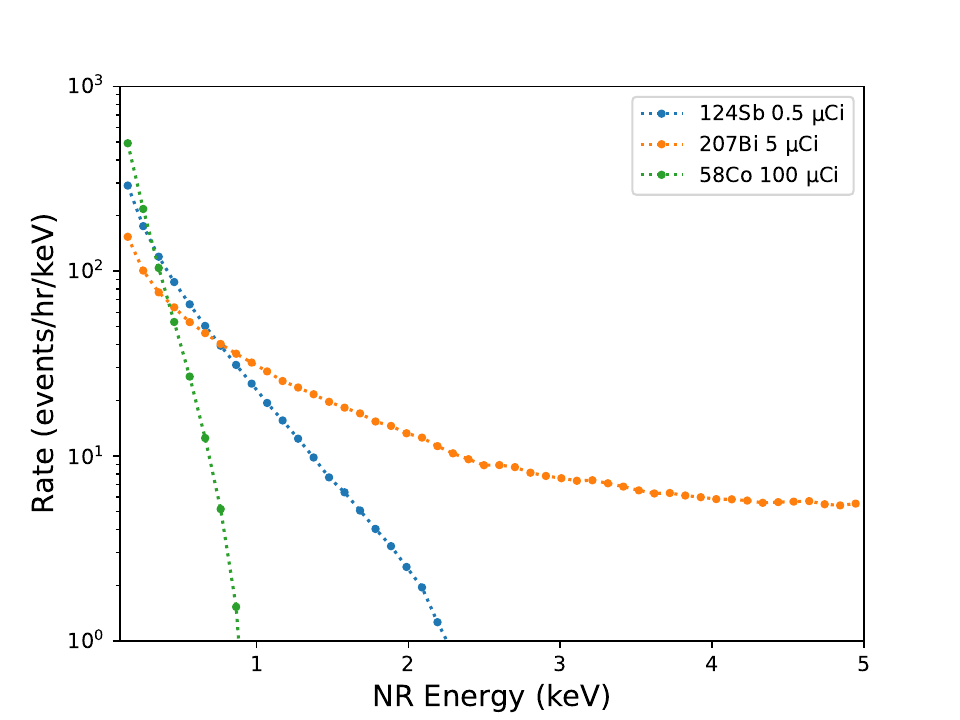}
    \caption{The simulated spectra of NRs from the photoneutron sources on argon in the detector. The simulated source is a gamma source in the center of a metallic beryllium cylinder. Source activities are chosen to give rates of around 50 bubbles/hour above a threshold of 80~eV.}
    \label{fig:PhotoneutronSpectra}
\end{figure*}

The $^{58}$Co-Be simulation contains 0.35 million neutron NRs, the $^{124}$Sb-Be simulation contains 0.58 million neutron NRs, and the $^{207}$Bi-Be simulation contains 0.71 million neutron NRs. The simulated NR spectra are shown in Figure~\ref{fig:PhotoneutronSpectra}. Due to the ER-blindness of bubble chambers, backgrounds from direct gamma rays are negligible, even with no gamma shielding of the sources.

\subsection{Thomson Scattering} \label{Thomson Scattering Sources}

Photon-nucleus elastic scattering includes nuclear Thomson scattering, Delbrück scattering, Rayleigh scattering, and nuclear resonance scattering \cite{AlanCoherentPhotons}. Rayleigh scattering is the dominant process, but is only significant for low-momentum transfer producing $<$30 eV NRs. For higher energy NRs that concern this calibration, nuclear Thomson scattering is dominant. As a shorthand, this paper refers to the combination of these interactions as Thomson scattering.

The NR energy ($E_{r}$) from Thomson scattering is given by:
\begin{equation}
    E_{r} = \frac{2p^{2}}{M} \sin^{2}\left(\frac{\theta}{2}\right),
\end{equation}
where $p$ is the gamma ray momentum, $\theta$ is the scattering angle, and $M$ is the mass of the nucleus. The high mass of the argon nucleus means that only a very small fraction of the gamma ray energy can be transferred by elastic scattering. The recoil energy is maximized when the scattering angle is 180$^{\circ}$, with maximum recoil energy:
\begin{equation}
    E_{r,max} = \frac{2p^{2}}{M}.
\end{equation}
Thomson scattering sources considered and their maximum NR energy are listed in Table \ref{table:simrates_T}.

\begin{table}
\caption{Information on the Thomson scattering sources considered for use in the calibration. The max NR column describes the maximum energy recoil on argon. The high energy $^{228}$Th gamma-ray comes from the $^{208}$Tl daughter isotope.}
\begin{tabular}{l l l l l}
\toprule
%Source Configuration & \multicolumn{2}{c}{ K-Shell Rate} & \multicolumn{2}{c}{L-Shell Rate} & \multicolumn{2}{c}{Energy Deposition Rate} \\
% & \multicolumn{2}{c}{[h$^{-1}$~ppm$^{-1}$]} & \multicolumn{2}{c}{[h$^{-1}$~ppm$^{-1}$]} & \multicolumn{2}{c}{[MeV~s$^{-1}$]} \\
Source & \multicolumn{1}{l}{Primary Gammas} & \multicolumn{1}{l}{Branching Ratio} & \multicolumn{1}{l}{Max NR Energy} & \multicolumn{1}{l}{Half-Life}\\
 & \multicolumn{1}{l}{[MeV]} & \multicolumn{1}{l}{[\%]} & \multicolumn{1}{l}{[eV]} & \multicolumn{1}{l}{[years]}\\
\hline
\hline
$^{22}$Na & 1.275 & 100 & 87 & 2.6\\
$^{60}$Co & 1.173 ,\ 1.333 & 100 ,\ 100 & 73,\ 95 & 5.3\\
$^{152}$Eu & 1.408 & 21 & 107 & 13.5\\
$^{88}$Y & 1.836 & 99 & 182 & 0.29\\
$^{207}$Bi & 1.064,\ 1.770 & 75,\ 7 & 61,\ 168 & 31.6\\
$^{124}$Sb & 1.691,\ 2.091 & 47 ,\ 5 & 152,\ 234 & 0.16\\
$^{228}$Th & 2.614 & 36 & 363 & 1.9 \\
\hline
\bottomrule
\end{tabular}
\label{table:simrates_T}
\end{table}

Thomson scattering simulations are done in Geant4 version 10.5.1, which includes the Japan Atomic Energy Agency (JAEA) elastic scattering package \cite{JAEA}. The source geometry in Geant4 is the same as for the photoneutron source simulations described in Section \ref{Photoneutron Scattering}, except without the $^{9}$Be cylinder. The JAEA code in version 10.5.1 had implementation errors, which were corrected. See Appendix \ref{JAEA changes} for details.

As with photoneutrons, directly simulating Thomson scattering is computationally intensive due to the low cross-section. For efficiency, Compton scattering is simulated instead, with the Thomson scattering rate then inferred from the Compton scattering rate. The ratio of the Thomson scattering cross-section to the Compton scattering cross-section gives the probability that a given Compton scatter would have instead been a Thomson scatter. This ratio, which depends on the energy of the gamma ray, is used as an $\mathcal{O}$(10$^{-5}$) weight factor. The Compton cross-section is calculated by integrating the differential cross-section from the Livermore Compton model over all energies \cite{Livermore}. The Thomson scattering cross-section is calculated by integrating the JAEA elastic scattering differential cross-section for >40 eV NRs. Each Compton scatter is randomly assigned an NR energy >40 eV with probability proportional to the differential cross-section. In testing, the bubble rates inferred from Compton scatters agree with directly simulated rates for each source. The Geant4 JAEA simulations were validated by comparing simulation results to existing cross-section data tables \cite{Kane,Falkenberg}.

\begin{figure}
    \centering
    \includegraphics[width=1.1\textwidth]{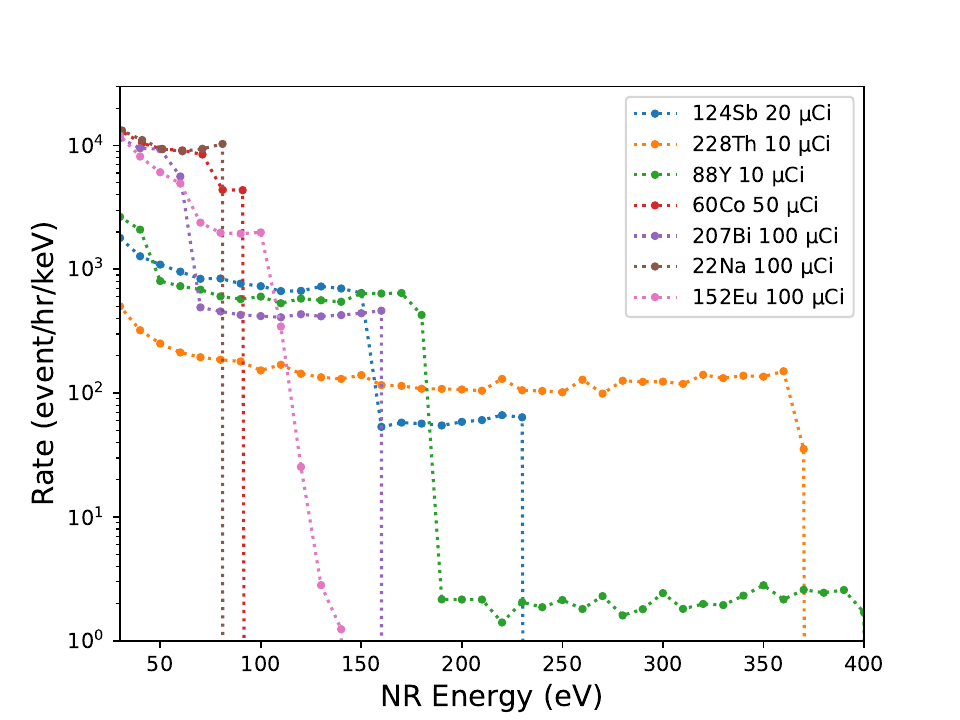}
    \caption{The simulated NR energy spectra from Thomson scattering sources on argon in the detector. The source activities were chosen to give rates of around 50 bubbles/hour above a threshold of 80~eV, but capped at 100 \textmu Ci. $^{228}$Th uses the highest activity easily available of 10 \textmu Ci.}
    \label{fig:ThomsonSpectra}
\end{figure}

The Thomson scattering sources considered cover a range of NR energies as shown in Figure~\ref{fig:ThomsonSpectra} and complement the higher energy NRs from the photoneutron sources. The simulations have Compton scattering sample sizes ranging from $^{152}$Eu with 0.11 million events to 1.8 million $^{60}$Co events. The subset of Thomson scattering sources used in the simulated calibrations presented in Section~\ref{Results} has at least 0.48 million events. These statistics are two orders of magnitude higher than expected bubble counts ($<$5000). In addition, rerunning a simulated calibration with new source simulations at these statistics does not change the results of any simulated calibration.

ER blindness is essential for the Thomson scattering calibration, because of the high rate of ERs from Compton scattering. Simulations of $^{228}$Th suggest ER rejection at the $10^{-8}$ level is necessary for ER backgrounds to be negligible. A similar rejection factor is already a requirement to control gamma ray backgrounds in rare event searches, and a high degree of ER rejection has been demonstrated in noble liquid bubble chambers~\cite{SBCxenonNRER}. 

\subsection{Thermal Neutron Capture} \label{Thermal Neutron Capture}

\begin{figure}[hbt!]
        \centering
        \includegraphics[height=5cm]{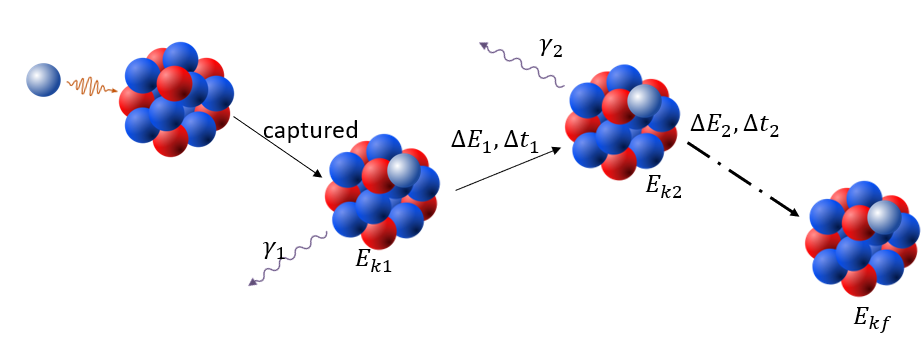}
        \caption{Argon nucleus recoil process after thermal neutron capture.} 
        \label{fig:TN_proc}
\end{figure}
Thermal neutrons have energies between \(10^{-2}\) eV and \(10^{-1}\) eV, and they are readily captured by many nuclei, including $^{40}$Ar and $^{36}$Ar. After capturing a thermal neutron, the resulting nucleus generally exists in an excited state, and it will emit a sequence of gamma rays as it relaxes to the new nuclear ground state. In SBC, the outgoing gamma rays will produce scintillation in the argon, flagging that a neutron capture has occurred. Each of the gamma rays will also provide a recoiling kick to the nucleus itself, with the energy of both the gamma rays and the recoil depending on the de-excitation path. Taking $^{40}$Ar for example:
\begin{equation}
    \prescript{40}{}{\mathrm{Ar}} + n \xrightarrow{}  \prescript{41}{}{\mathrm{Ar}} + \sum_{i=1}^{n} \gamma_{n},
\end{equation}
where the total number $n$ and the sequence of gamma energies is determined by the de-excitation path.

For the case where $n=1$, the single gamma ray would result in a monoenergetic NR with energy 
\begin{equation}
    E_r = \frac{E_\gamma^2}{2mc^2}
\end{equation}
where $m$ is the mass of the recoiling nucleus. In cases where there is a sequence of gamma rays, the process is more complicated, depending on the stopping time of the recoiling nucleus and the angular distribution of the gamma ray sequence. For example, if a nucleus emits two nearly simultaneous gamma rays of similar energy in opposite directions, there would be almost no induced NR energy. On the other hand, if the second gamma ray emission occurs after the recoil induced by the first has completely stopped, the total amount of recoil energy would be the full sum of each individual recoil. 

A diagram of the post-decay gamma ray sequence is shown in Figure \ref{fig:TN_proc}. The argon nucleus captures a thermal neutron and emits $\gamma_{1}$, gaining initial kinetic energy $E_{k1}$. The recoiling nucleus then deposits a portion of that kinetic energy $\Delta E_1$ into the surrounding argon liquid in a time $\Delta t_1$ before the next emission of $\gamma_{2}$. The recoil momentum vector from the second gamma adds to the residual momentum vector from the first recoil, creating a nucleus with kinetic energy $E_{k2}$, which again deposits $\Delta E_2$ in $\Delta {t_2}$ before the third gamma emission. The nucleus gains momentum and deposits kinetic energy into the liquid in this fashion until the last gamma is emitted, after which the nucleus deposits all remaining kinetic energy, $E_{kf}$. The total NR energy deposited into the liquid is:

\begin{equation}
\label{eq:TN_E_r}
    E_r = \sum_{i=1}^{n-1} \Delta E_{i} + E_{kf}.
\end{equation}
To model this process, a custom simulation including known energy levels, lifetimes, and gamma emission branching fractions of the nuclear de-excitations of $^{41}$Ar and $^{37}$Ar following thermal neutron capture on $^{40}$Ar and $^{36}$Ar is developed. To model the stopping of the argon nuclei in the liquid, Lindhard theory is utilized, cross-checked using the Stopping and Range of Ions in Matter (SRIM) package \cite{Ziegler2010}. More details of the model are discussed in Appendix \ref{Stopping of keV Argon Recoils}.

\begin{figure}[hbt!]
        \centering
        \includegraphics[height=6cm]{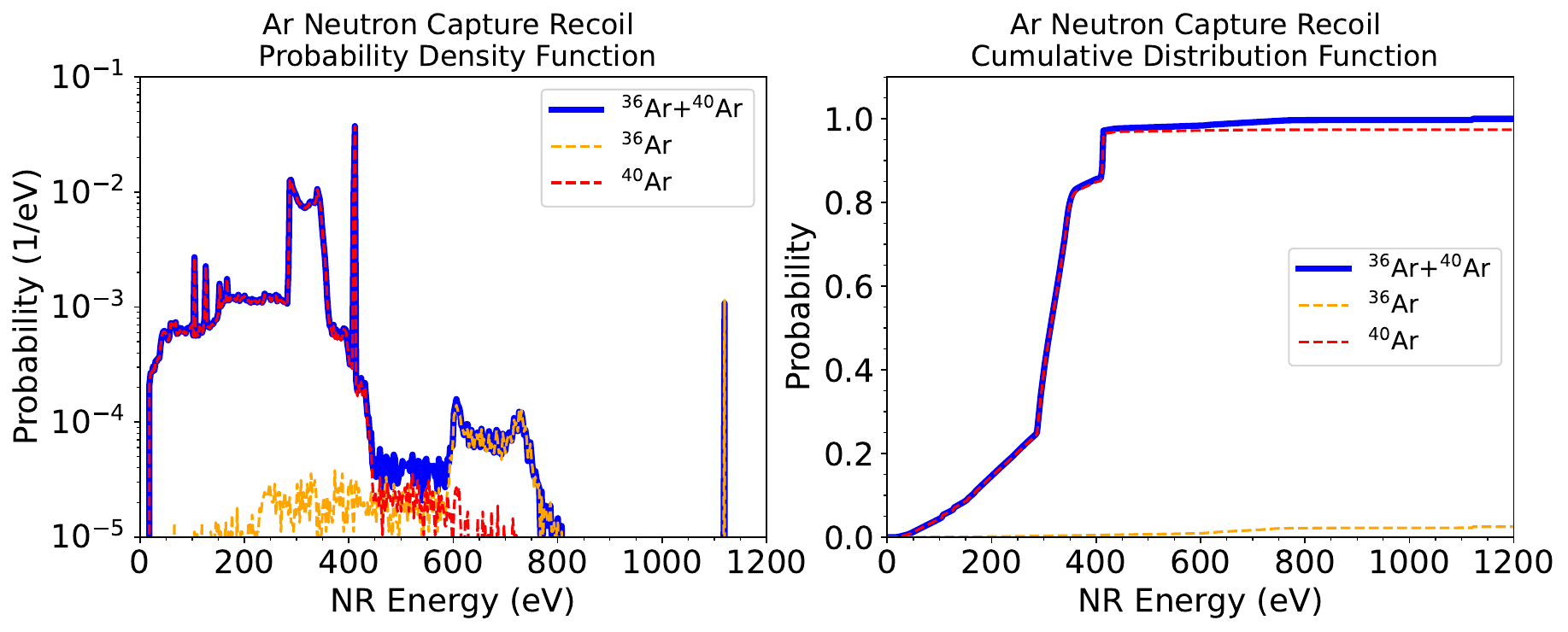}
        \caption{Probability density function and cumulative distribution function for the NR energies of $^{41}$Ar and $^{37}$Ar nuclei following thermal neutron capture on $^{40}$Ar and $^{36}$Ar labeled by the initial state. The relative contributions of $^{40}$Ar and $^{36}$Ar are determined from the natural abundances and neutron capture cross sections shown in Table~\ref{tab:Ar_norm}.}
        \label{fig:TN_spectrum}
\end{figure}

The final predicted NR spectrum is simulated with 0.2 million events and is given in Figure~\ref{fig:TN_spectrum}. The range of energy deposition below 600 eV comes from capture on $^{40}$Ar, whose multi-gamma sequences are dominated by one of a few high-energy gamma lines. The high energy line above 1100 eV is from capture on $^{36}$Ar, which de-excites primarily through a single gamma ray of 8.79~MeV. The relative normalization comes from the natural abundances and neutron capture cross sections, which are shown in Table \ref{tab:Ar_norm}~\cite{Durnford2018}. The $^{38}$Ar contribution is not included both because of its small overall contribution and lack of experimental data from characterizing its de-excitation scheme.

\begin{table}[]
\caption{Thermal cross sections (in milli-barns at 2200 m/s incident velocity), isotopic abundances, and the overall contribution of each isotope included in the simulation of neutron
capture in argon.}
    \centering
    \begin{tabular}{ llll }
 \hline
 Isotope & $\sigma_{n,\gamma}$  &Abundance& Contribution \\
  &[mb] &\%&\% \\
 \hline
  \hline
 $^{36}$Ar   & 5200   &0.3&   2.5\\
 $^{38}$Ar&   800  & 0.1  &0.1 \\
 $^{40}$Ar&   700  & 99.6  &97.4 \\
 \hline
 \bottomrule
    \end{tabular}
    \label{tab:Ar_norm}
\end{table}

\subsubsection{Thermal Neutron Source Design}
The primary background to the thermal neutron source calibration is scattering of fast neutrons (both elastic and inelastic) produced by the source itself. Even heavy moderation of standard $^{241}$Am-Be and $^{252}$Cf sources preserves a significant fraction of fast neutrons. Therefore, this study utilizes the lower energy $^{241}$Am-Li source recently produced at the University of Alabama for the LZ collaboration as a baseline~\cite{LZ_AmLi}. Higher rate $^{241}$Am-Li sources have previously been discussed in the literature~\cite{PNNL_AmLi} and would also work with some moderation. Photoneutron sources are a third alternative under study. 

\begin{figure*}
    \centering
    \includegraphics[width=0.8\textwidth]{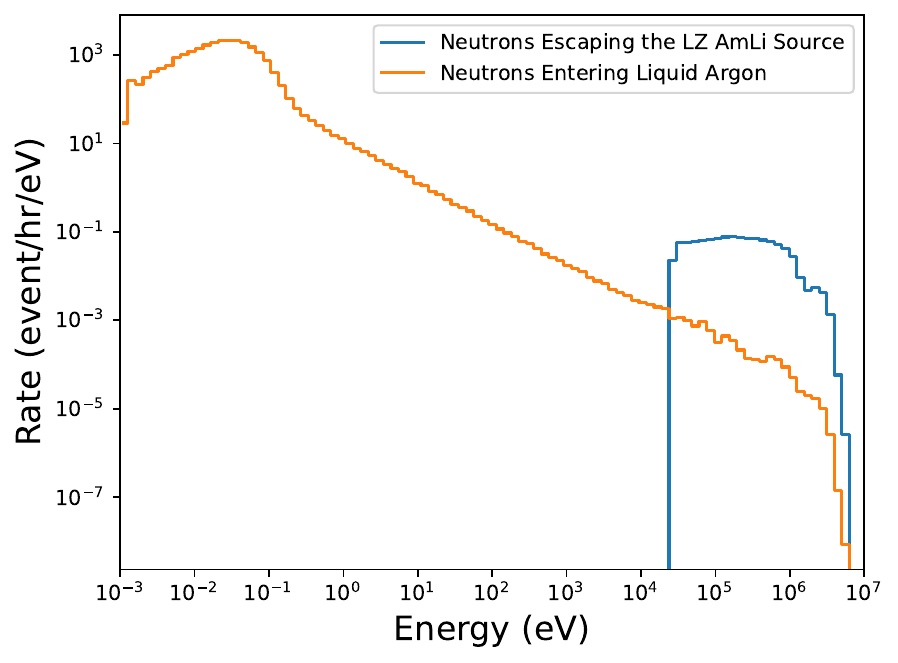}
    \caption{Energy spectrum of neutrons escaping from the AmLi source (assuming the combination of AmLi-2 + AmLi-3 from Table~\ref{tab:AmLi_config}) and upon entering the liquid argon volume. }
    \label{fig:AmLi_spectrum}
\end{figure*}

$^{241}$Am-Li is an ($\alpha$, n)-type source. The main channel that produces neutrons is:
\begin{equation}
\label{eq:AmLi_1}
    \prescript{241}{}{\mathrm{Am}} \xrightarrow{}  \prescript{237}{}{\mathrm{Np}} +  \alpha,
\end{equation}
\begin{equation}
\label{eq:AmLi_2}
    \alpha + \prescript{7}{}{\mathrm{Li}}\xrightarrow{} \prescript{11}{}{\mathrm{B}}^* ,
\end{equation}
\begin{equation}
\label{eq:AmLi_3}
    \prescript{11}{}{\mathrm{B}}^*\xrightarrow{} \prescript{10}{}{\mathrm{B}}^*  +n .
\end{equation}
With a pure Li target, the maximum neutron energy is 1.5~MeV. There are 3 $^{241}$Am-Li sources described in Ref.~\cite{LZ_AmLi} and their neutron and gamma yields are shown in Table ~\ref{tab:AmLi_config}.
\begin{table}[]
\caption{LZ $^{241}$Am-Li sources configurations~\cite{LZ_AmLi}.}
    \centering
    \begin{tabular}{ lllll}
 \hline
 Source &  Neutron Emission Rate &$\gamma$ Emission Rate \\
  &[Hz] &[Hz]& \\
 \hline
  \hline
 AmLi-1   & 18$\pm$2   &368$\pm$59\\
 AmLi-2&   9$\pm$1  & 239$\pm$38   \\
 AmLi-3&   12$\pm$1  & 318$\pm$51   \\
 \hline
 \bottomrule
    \end{tabular}
    \label{tab:AmLi_config}
\end{table}
The as-built sources potentially include some chemical impurities leading to a higher energy neutron tail, and this work assumes the energy spectrum of $^{241}$Am-Li contaminated with (NO$_3$)$_3$ from~\cite{LZ_AmLi}, shown in Fig.~\ref{fig:AmLi_spectrum}.  To achieve the desired rates of approximately 50 bubbles/hour, the AmLi-2 and AmLi-3 sources are placed 71 cm from the SBC detector (center to center distance), with thermalization achieved by the SBC detector itself, particularly the HDPE castle. The energy spectrum of neutrons entering the liquid argon is also shown in Figure~\ref{fig:AmLi_spectrum}.

Because of the (NO$_3$)$_3$ contamination, fast neutrons from the source remain the dominant background over external backgrounds. A good thermal neutron capture event is characterized by the NR (forming a bubble) along with scintillation light produced by the gamma rays interacting in the argon. By tagging with the scintillation light, low energy neutron elastic scattering without observable scintillation photons, labeled as ``bubble-only background,'' can be distinguished and eliminated from neutron capture with high efficiency. However, background neutrons from the source can still mimic the signal in two ways: ``Correlated ER backgrounds'' include fast neutrons producing a bubble via elastic or inelastic scattering in the argon, followed by scintillation produced by the inelastic scatter directly or by scattering or capture elsewhere in the geometry leading to gamma emission that travels back to the liquid argon; ``Hard scatter backgrounds'' arise when the bubble-inducing NR itself produces enough scintillation light to be visible. 

\begin{figure*}
    \centering
    \includegraphics[width=0.8\textwidth]{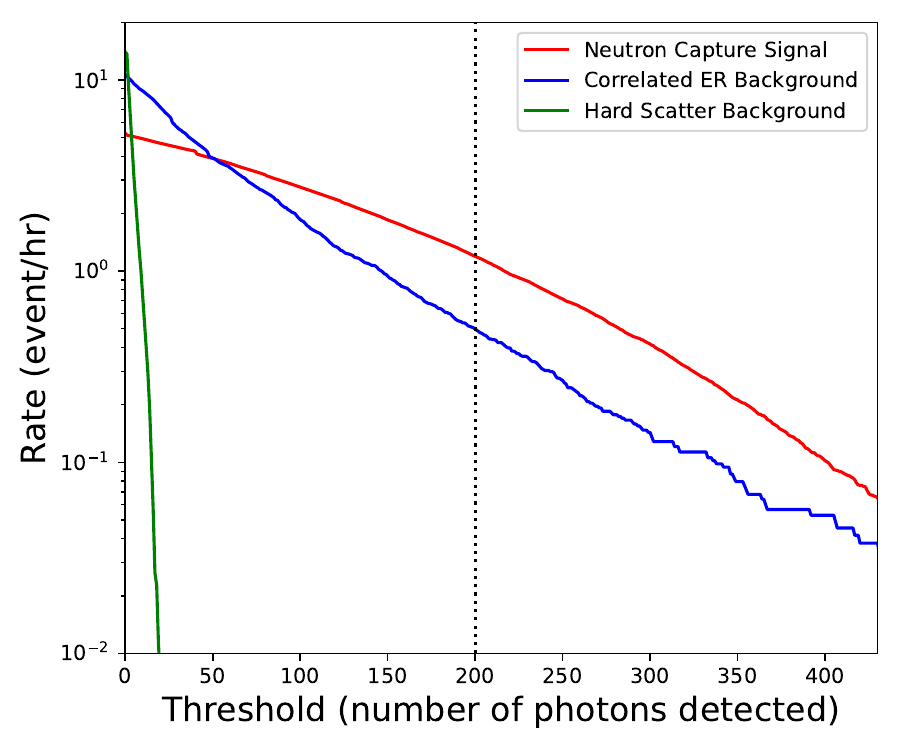}
    \caption{Signal and background bubble rates as a function of threshold in the number of scintillation photons detected assuming the nucleation efficiency curve is a normal cumulative distribution function with a width of 50 eV centered at 400 eV. Light yields of 10(40) photons/keV for NR(ER) are assumed, along with a photon detection efficiency of 0.6\%. The hard scatter background is eliminated by a modest threshold in the light signal. As an example, a threshold of 200 detected photons (black dotted vertical line) achieves a signal to correlated ER background ratio of 2.41 while preserving an adequate signal rate. }
    \label{fig:Cf_SN}
\end{figure*}

All classes of of neutron correlated backgrounds can be mitigated by placing a cut on the size of the scintillation light signal. Figure \ref{fig:Cf_SN} shows how the neutron capture signal and background rates change with the number of photons detected in an event, assuming light yields of 10(40) photons/keV for NR(ER) and a photon detection efficiency of 0.6\%.  At a threshold of approximately 200 detected photons, the $^{241}$Am-Li sources are found to provide a good calibration signal of nearly monoenergetic NRs around 400~eV, with a projected signal rate of 1.18 bubbles per hour, a post-veto background rate of 0.49 bubbles per hour, and approximately 44.71 vetoed bubbles per hour.

Uncorrelated backgrounds are produced when accidental coincidences of background bubbles and scintillation light occur in a window of roughly 50 \textmu s (the resolution for measuring the time of bubble formation). Given the low ambient bubble background and the scintillation energy threshold for the capture tag, these are deemed to be negligible compared to the correlated rates. Table \ref{tab:TN_veto} shows the capture signal and background rate before and after vetoing. Operationally, the total bubble rate without gamma tagging should be limited to approximately 50 bubbles/hour and the current simulated configuration achieves 50.46 bubbles/hour.

\begin{table}[]
\caption{Neutron capture tagging information at 400 eV NR energy and tagging threshold of 200 photons detected.}
    \centering
    \begin{tabular}{ llllll }
 \hline
 Bubbles/Hour & Neutron Capture &Correlated ER& Hard Scatter& Bubble Only &Total \\
  &Signal &Background&Background&Background & Background \\
 \hline
  \hline
 Before Tagging   & 5.26   &10.69&   14.05&20.56&45.20\\
 After Tagging&   1.18  & 0.49  &0.00 &0.00&0.49\\
 \hline
 \bottomrule
    \end{tabular}
    \label{tab:TN_veto}
\end{table}

\section{Simulated Calibrations} \label{Simulated Calibration}

In this section, simulated calibrations for the SBC-LAr10 experiment are described, making use of the source simulations from Section~\ref{Source Simulations}. These simulated calibrations inform the choice of sources to deploy and the precision with which the nucleation efficiency function can be measured. %Simulated calibrations are run for various calibration schemes, varying the source list, constraints on systematic uncertainties, and energy thresholds. 
The calibration source list in each simulated calibration contains a subset of the sources presented in Section~\ref{Source Simulations}. The impact of systematic uncertainties such as source strength uncertainty and simulation uncertainty are incorporated. The source activities are chosen for a target rate of 50 bubbles/hour, roughly the highest rate the chamber can effectively operate at, and far higher than the expected external neutron background of $\sim$1.4 events per hour (see Table~\ref{table:systematics}).  For each source, 100 hours of livetime is assumed, such that systematic uncertainties are larger than statistical uncertainties. Calibrations are simulated for efficiency functions centered from 80 eV to 400 eV. %Previous $\sim$keV NR calibrations suggest effective nucleation occurs at roughly double the Seitz threshold in C$_{3}$F$_{8}$ \cite{pico60c3f8}. While lower effective thresholds are possible for sub-keV argon Seitz thresholds, the 40 eV minimum Seitz threshold will likely correspond to roughly 80 eV.

%\begin{table}
%\caption{The background bubble rate simulated for the calibration site based on Geant4 simulations of rock and muon induced neutrons for the MINOS tunnel at Fermilab, along with statistical uncertainties. Background simulations all assume a step function threshold at 100~eV. Because of large systematic uncertainties on backgrounds, a more conservative background assumption is also considered, approximately three times the simulated rate.}
%\begin{center}
%\begin{tabular}{  l l l  }
%\hline
% Multiplicity &  Simulated Background& Conservative Background\\ 
% (bubbles/event) &  (events/hour)& (events/hour) %\\
%\hline
%\hline
% 1&  1.41 $\pm$ 0.05&5.00\\ 
% 2&  0.45 $\pm$ 0.03&1.59\\ 
% 3&  0.20 $\pm$ 0.02&0.69\\ 
% 4&  0.10 $\pm$ 0.01&0.37\\ 
% 5&  0.05 $\pm$ 0.01&0.16\\ 
% 6&  0.03 $\pm$ 0.01&0.11\\ 
% 7&  0.02 $\pm$ 0.01&0.08\\  
% 8&  0.01 $\pm$ 0.01&0.02\\ 
% 9+& 0.02 $\pm$ 0.01&0.06\\
%\hline
%\bottomrule
%\end{tabular}
%\label{table:backgroundMult}
%\end{center}
%\end{table}

\subsection{Mock Datasets} \label{Toy Datasets}
\begin{figure}
    \centering
    \includegraphics[width=0.8\textwidth]{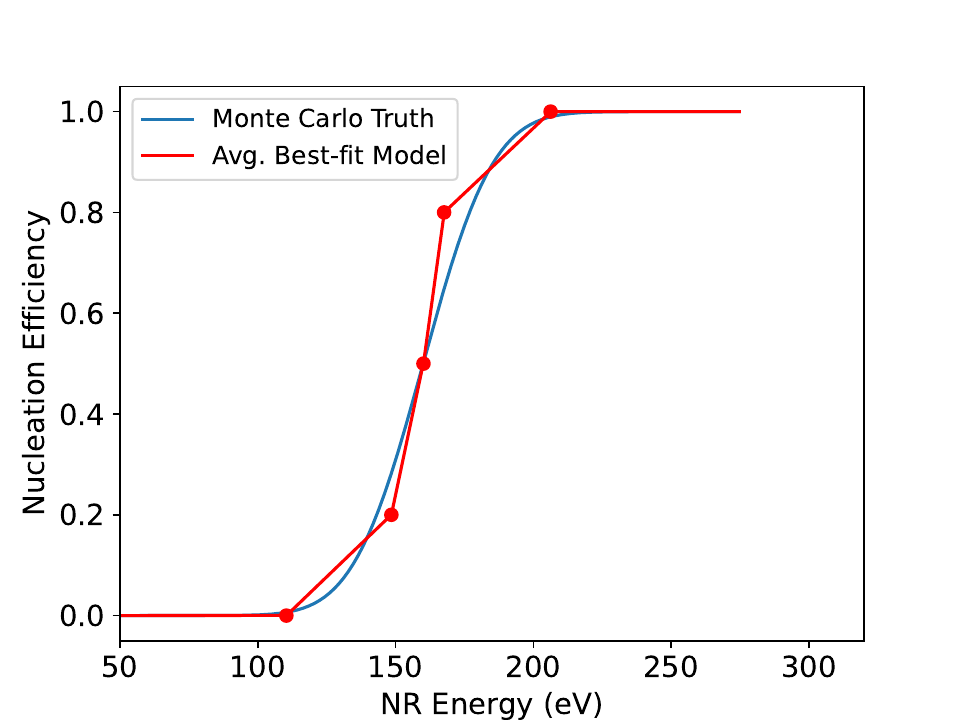}
    \caption{Example of the NCDF used as the Monte Carlo true nucleation efficiency function shown in blue. This function has a mean of 160 eV and a Gaussian width of 20 eV. An example model piecewise nucleation efficiency function as explained in Section \ref{Fitting Methodology} is shown in red.}
    \label{fig:efficiency-function}
\end{figure}

The calibration source simulations are used to create mock datasets that mimic data in a real calibration. To get a rate of bubbles from a simulated NR spectrum, a Monte Carlo true probability of bubble formation, or nucleation efficiency function, must be chosen. For the purposes of these mock calibrations, a Normal Cumulative Distribution Function (NCDF) is used as the Monte Carlo true efficiency function, because it roughly models the shape of the efficiency functions observed in PICO Freon calibrations \cite{pico_calibration}. The NCDF has the form:
\begin{equation}
   \epsilon_{\text{true}}(E_{r}) = \frac{1}{\sqrt{2 \pi} \sigma} \int_{-\infty}^{E_{r}} e^{-\frac{1}{2}[(E-T) / \sigma]^{2}}dE,
\end{equation}
where $\epsilon_{\text{true}}$ is the true probability of nucleating a bubble, $\sigma$ is the Gaussian width, $T$ is the Gaussian mean (in this case the 50\% nucleation efficiency point), and $E_{r}$ is the NR energy. An example of this function is shown in blue in Figure \ref{fig:efficiency-function}.

The statistics of interest for the mock datasets are the total number of bubbles and the number of events with one or more bubbles. These parameters are chosen to provide a balance between computation time and preserving some information on the bubble multiplicity for the photoneutron sources. For each source, the expected total number of bubbles $\langle B \rangle$ is calculated by binning the simulated recoils in energy:
\begin{equation}
    \langle B \rangle = \frac{t_{\text{cal}}}{t_{\text{sim}}} \sum_{i=1}^{N_{\text{bins}}} \epsilon_{\text{true}}(E_{i})W(E_{i}) + \text{background},
\end{equation}
where $\epsilon_{\text{true}} (E_{i}$) is the nucleation efficiency for a given energy bin $E_{i}$, $W(E_{i})$ is the NR count for that energy bin, $t_{\text{cal}}$ is the planned calibration live time (100 hours for each source), and $t_{\text{sim}}$ is the simulated live time.  The bin size for Thomson scattering is 1 eV, and for photoneutron sources the bin size is 0.5 eV below 500 eV and 2 eV above 500 eV. The thermal neutron capture simulations are unbinned. The expected total number of events $\langle V \rangle$ (all events with at least one bubble) is also calculated. For Thomson scattering and neutron capture, $\langle V \rangle=\langle B \rangle$. For the photoneutron sources, some events have multiple recoils. Each simulated recoil is randomly assigned as a bubble or not with a probability according to the nucleation efficiency, and the number of events with $\geq$1~bubble is counted. 

For external background, both the simulated rate of 1.41 single bubble events per hour and a more conservative assumption of 5 events per hour are used for Thomson scattering and photoneutron sources (see Table~\ref{table:systematics}). Multiple bubble backgrounds are also considered for photoneutron sources. An average rate of 0.49 single bubble events per hour is assumed for the thermal neutron source background after scintillation tagging (see Table~\ref{tab:TN_veto}), assumed to be constant with threshold since it is dominated by recoils well above threshold. Prior to NR calibration, the detector threshold will be increased until ER backgrounds are negligible.

The Monte Carlo expected counts form the basis of the mock datasets, one number per Thomson scattering or thermal neutron capture source (the total number of bubble events), and two numbers for the photoneutron sources (total number of bubbles and total number of events). Two types of systematic uncertainty are considered: source strength uncertainty and simulation uncertainty. The simulation uncertainty models uncertainty in source geometry as well as simulated cross-sections, and is common for all sources of a given type (photoneutron, Thomson scattering, or thermal neutron capture).  To implement each uncertainty, the rate for each source is multiplied by a random sample from a log-normal distribution centered on one. Uncertainty in the background rate is negligible since the backgrounds can be measured in situ --- the external background can be measured with no sources, and backgrounds from the thermal neutron source can be measured by raising the bubble nucleation threshold.

At each detection threshold considered, 50 mock datasets are generated. Each mock dataset is generated independently from other datasets with different Poisson samples for background and signal, and different random values for the systematic uncertainties in each source and calibration method. A mock dataset includes data from all sources used at that threshold. The fit to each mock dataset represents what a real calibration campaign at a given threshold could look like, and the spread of the fits informs the expected uncertainty in the real calibration.

\subsection{Fitting Methodology} \label{Fitting Methodology}
The mock datasets are fit to a piecewise nucleation efficiency function, which avoids making assumptions about the shape of the true efficiency function. This is the same function used by PICO to fit their calibration data, and an example of this functional form is shown in red in Figure~\ref{fig:efficiency-function} alongside the Monte Carlo true efficiency function. The function has five node energies that are fixed at 0, 0.2, 0.5, 0.8, and 1 on the efficiency axis but are free to move on the NR energy axis. For example, the 0.2 node represents the energy at which a NR forms a bubble 20\% of the time. The function linearly connects the efficiency nodes. The NR energy values associated with each node are fit to best match the calibration data, with the constraint that the nodes are forced to monotonically increase.

The model has five parameters for the nucleation efficiency function (one for each of the monotonically increasing efficiency nodes), a nuisance parameter for the activity of each source, and a simulation nuisance parameter for each type of source. For each set of parameters, the model generates a number of expected bubbles and events based on the simulated NR spectrum and the nucleation efficiency function. The model count includes expected background.

For the model, a slight change is made in photoneutron event counting compared to the mock datasets. The assumption is made that if the highest energy recoil in an event does not make a bubble, then none of the recoils in that event do. Therefore, the maximum energy NR in each event $E_{i,\text{max}}$ are the only NRs used to generate the event rate,  allowing the events to be binned in $E_{i,\text{max}}$, reducing computation time by orders of magnitude.  With this assumption, the model expected number of photoneutron events $v$ is given by:
%\begin{equation}
%   \langle r_{V} \rangle = \frac{1} {t_{\text{sim}}} \sum_{i=1}^{N_{\text{events}}} \epsilon_{\text{true}}(E_{i,\text{max}})W(E_{i}).
%\end{equation}
\begin{equation}
   v = \frac{t_{\text{cal}}} {t_{\text{sim}}} \sum_{i=1}^{N_{\text{events}}} \epsilon_{\text{true}}(E_{i,\text{max}})W(E_{i}) + \text{background}.
\end{equation}
This assumption changes the event rate by less than 0.3\% and simulations show no significant change in the fitting result.

The chi-squared is calculated with the Pearson's chi-squared test~\cite{Pearson}:
\begin{equation}
    \chi^{2} = \sum_{i=1}^{n} \left( \frac{[V_{\text{i}}-v_{\text{i}}(E_{0},...E_{1},a_{a,\text{i}},a_{c,\text{i}})]^{2}}{V_{\text{i}}} +
    \frac{\left[B_{\text{i}}-b_{\text{i}}\left(E_{0},...E_{1},a_{a,\text{i}},a_{c,\text{i}}\right)\frac{V_{\text{i}}}{v_{\text{i}}}\right]^{2}}{B_{\text{i}}} \right)+\sum_{j=1}^{n+2} \frac{a_{j}^{2}}{\sigma_{j}},
\end{equation}
where $n$ is the number of sources, $E_0...E_1$ are the energies of a given efficiency node, $a_j$ are the nuisance parameters (which are subdivided into $a_{a}$, the activity for a given source, and $a_{c}$, the simulation nuisance parameter common for a given source type), and $\sigma_j$ represents one standard deviation for that nuisance parameter. For each source, the event count is given as $V$ for the mock dataset and $v$ for the model's fit value. For photoneutron sources, the number of bubbles in the dataset is given as $B$ and the number of bubbles in the model is given as $b$. Since statistical fluctuations from the model in the number of events and number of bubbles are highly correlated, the model number of bubbles is normalized to the mock dataset event count ($b_{\text{i}} \rightarrow b_{\text{i}}V_{\text{i}}/v_{\text{i}}$), eliminating the correlation. Note that the second term can also be written in terms of the average multiplicity of mock dataset ($M_{\text{i}}=B_{\text{i}}/V_{\text{i}}$) and model ($m_{\text{i}}=b_{\text{i}}/v_{\text{i}}$):

\begin{equation}
    \frac{\left[B_{\text{i}}-b_{\text{i}}\frac{V_{\text{i}}}{v_{\text{i}}}\right]^{2}}{B_{\text{i}}} 
    =
    \frac{\left[M_{\text{i}}-m_{\text{i}}\right]^{2}}{M_{\text{i}}} V_{\text{i}}.
\end{equation}
For Thomson and neutron capture sources the multiplicity is always 1 and that term is ignored ($M$ - $m$ = 0).

The chi-squared is minimized using a Markov chain Monte Carlo \cite{Markov}. Small iterative changes, or "steps" are made to the model parameters, and over a large number of iterations, the parameters that result in the lowest chi-squared are taken as the best fit parameters. For each step, a uniformly sampled random value is added to each model parameter. This value can be positive or negative and represents a small change to the parameter value. The maximum step size for the efficiency nodes is 0.5 eV. The maximum step size for the nuisance parameters is a 0.1\% change to the simulation nuisance parameter and a 0.5\% change to the source activity parameter.

%B$ is the dataset mean multiplicity (bubbles/event) and $m$ is the model fit mean multiplicity. The $\chi^{2}$ term from the multiplicity is multiplied by true event rate so it as the same impact on the model. For Thomson sources the multiplicity is always 1 and that term is ignored ($M$ - $m$ = 0).

The fit uses the Hasting-Metropolis algorithm to accept or reject new steps \cite{HastingsMetro}. If the chi-squared calculated at the next step is lower than the previous step, the model takes the new values. If the chi-squared increases, the model most likely reverts to the old parameters, but has a chance to keep the worse value to avoid being stuck in a local minimum. The probability of keeping the new values decreases as chi-squared increases following this equation:
\begin{equation}
    P = \frac{\chi_{old}^{2}}{\chi_{old}^{2}+\chi_{new}^{2}}.
\end{equation}
This is repeated in a chain of steps, a “walker,” that explores the lowest chi-squared regions of parameter space.

Each fit involves 2 walkers with a different set of random initial parameters, and each walker runs for hundreds of thousands of steps or more. The starting set of parameters are kept between 50 and 230 eV with a randomized increase in energy between each parameter. This process reduces the impact of outliers on the fit. The initial guess for the nuisance parameters is to assume the systematic errors are zero. Using more than 2 walkers did not provide additional benefit since outliers were rare. 

\subsection{Calibration Assessment}
Calibration uncertainty is assessed in two ways: firstly, by comparing the effective threshold of the best fit function to the Monte Carlo truth, and secondly by comparing expected bubble rates from a dark matter or CE$\nu$NS signal. The effective threshold is calculated as the NR energy at which a Heaviside step function would turn on in order to have the same integrated area. The NCDF is centered on its effective threshold, but the effective threshold of the piecewise best fit function is not necessarily at the 50\% efficiency node and is calculated as follows:
\begin{equation}
    T_{\text{effective}} = E_{max} - \int_{0}^{E_{max}}{f(E)dE},
\end{equation}
where $T_{\text{effective}}$ is the effective threshold, $E$ is recoil energy, $f(E)$ is the value of the nucleation efficiency function for that energy, and $E_{max}$ is the energy that all models and the Monte Carlo true efficiency functions have reached 100\%, which occurs by 5~keV.

To compare the signal bubble rates the percent difference between model and true bubble rates for dark matter or reactor neutrinos is calculated based on expected NR spectra. Dark matter of mass 2~GeV$/c^{2}$ was chosen for this comparison. 
\section{Simulated Calibration Results}
\label{Results}
The threshold precision achieved by various calibration campaign schemes is investigated with the simulated calibrations. These schemes differ in the set of calibration sources chosen, the threshold at which the calibration takes place, constraints on systematic uncertainties, and background assumptions. The calibration is simulated at four energy thresholds, 80, 160, 320, and 400~eV, where the threshold is the 50\% efficiency point of the NCDF, and the Gaussian width scales with threshold (10, 20, 40, and 50~eV, respectively).
\begin{table}
\caption{Systematic uncertainties and backgrounds associated with standard precision and high precision simulated calibrations. Source uncertainties are uncorrelated between sources. Simulation uncertainties are common mode within each source type. External background (singles/hour) applies to the Thomson calibrations. External background (events/hour) and (bubbles/hour) applies to the photoneutron calibrations, which include multiple bubble events. The scintillation tag will remove nearly all external backgrounds for the neutron capture calibration. The high precision external background rates are from Geant4 simulations of rock and muon induced neutrons for the MINOS tunnel at Fermilab, assuming a step function threshold at 100~eV. For standard precision, the external background is conservatively assumed to be 3.5 times higher than simulated.}
\begin{center}
\begin{tabular}{  l l l  }
\hline
 &  Standard precision& High precision\\
\hline
\hline
 Source uncertainties&  5\%&1\%\\ 
 Thomson sim. uncertainty&  10\%&2\%\\ 
 Photoneutron sim. uncertainty&  20\%&5\%\\
 Neutron capture sim. uncertainty& 20\%&5\%\\ 
 External background (singles/hour) &  5.00&1.41\\
 External background (events/hour) &  8.12 &2.29\\
External background (bubbles/hour) &  14.68 &4.14 \\
 Neutron capture background (singles/hour) & 0.49 &0.49\\
 %Background& &\\
 \hline
 \end{tabular}

%\begin{tabular}{  c  lc  l}
% Standard Precision&  Thomson& Photoneutron&Neutron Capture\\
%\hline
%\hline
% Source Strength&  5\%&5\%&5\%\\ 
% Common Mode&  10\%&20\%&20\%\\ 
% Background&  5.00 singles/hour&5.00 singles/hour&0.49 singles/hour\\
% \hline
% High Precision& Thomson&Photoneutron&Neutron Capture\\ 
% \hline
%\hline
% Source Strength& 1\%&1\%&1\%\\
% Common Mode& 2\%& 5\%&5\%\\
% Background& 1.41 singles/hour& 1.41 singles/hour&0.49 singles/hour\\
% \hline
% \bottomrule
% \end{tabular}

\label{table:systematics}
\end{center}
\end{table}

The calibration requirements for a CE$\nu$NS measurement (5\% threshold uncertainty) are more stringent than for a dark matter search (20\% threshold uncertainty), thus two sets of calibrations are presented in Table \ref{table:systematics}, differing in the assumed systematic uncertainties and backgrounds. Eckert and Ziegler~\cite{EckertZiegler} calibrates sources supplied to SBC with a 3\% to 5\% uncertainty depending on the isotope, giving a baseline source strength uncertainty. For the standard precision dark matter search calibration, SBC is confident that the Thomson (10\%) and neutron (20\%) systematic uncertainties can be achieved in the near-term, based on previous calibrations~\cite{SBCxenonNR,pico_calibration}. A more aggressive set of systematic constraints is assumed for the high precision calibration. Uncertainty in source activity and source geometry could be constrained by a dedicated study of scintillation light collection from calibration sources, allowing future calibration efforts to achieve these lower systematic uncertainties. Simulated calibrations suggest that achieving a 5\% or lower uncertainty in threshold requires 5\% or lower systematic uncertainties. The precision calibration assumes the simulated external background rate, rather than the more conservative assumption of the standard precision calibration. A modest amount of hydrogenated shielding can be added if the background rate is higher than expected.

A source list that performs well across a broad range of energies includes 5 \textmu Ci $^{207}$Bi-Be and 0.5 \textmu Ci $^{124}$Sb-Be photoneutron sources, and 10 \textmu Ci $^{228}$Th, 20 \textmu Ci $^{124}$Sb, and 50 \textmu Ci $^{60}$Co Thomson scattering sources. The low energy neutrons from $^{58}$Co-Be would be useful, but the low rate of neutron generation make the source impractical. A 10 \textmu Ci $^{88}$Y Thomson scattering source also performs well, especially for 100--200 eV thresholds, and could be added as an additional calibration source. The thermal neutron calibration assumes the LZ $^{241}$Am-Li sources are available for loan. The three lower energy thresholds are calibrated with Thomson scattering and photoneutron sources. The 400~eV threshold is above the maximum recoil energy for all the Thomson scattering sources, but the thermal neutron source is well suited to this energy, thus the calibration uses thermal neutron and photoneutron sources.

\begin{figure}
    \centering
    \includegraphics[width=0.8\textwidth]{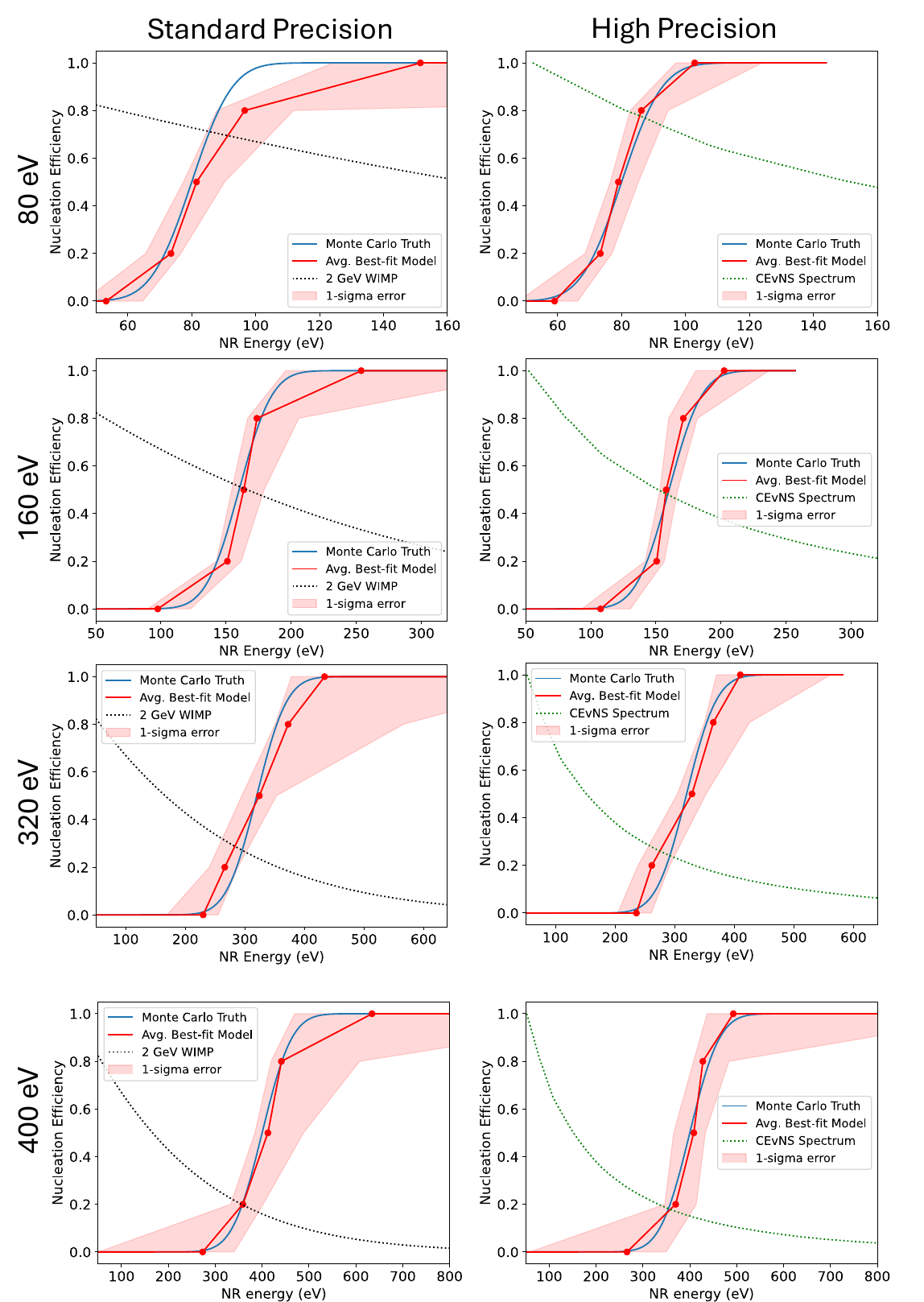}
    \caption{The best fit functions for six simulated calibrations compared to the Monte Carlo true efficiency functions. Blue lines represent the true efficiency functions centered at 80, 160, 320, and 400 eV with Gaussian widths of 10, 20, 40, and 50 eV, respectively. At the 400 eV threshold a thermal neutron capture and photoneutron calibration is performed, and for the three lower energy thresholds, a Thomson scattering and photoneutron calibration is used. The red line represents the median best fit mock dataset, and the red band represents 1-sigma spread in mock dataset best fits. Left column: standard precision calibration results with the set of background and systematic assumptions associated with near-term calibration for a dark matter search. A normalized NR spectrum from 2 GeV/$c^{2}$ dark matter in argon is shown in black. Right column: high precision simulated calibration results that assume the backgrounds and systematics that could be attained by a calibration campaign to support a future reactor CE$\nu$NS experiment. A normalized reactor CE$\nu$NS spectra is shown in green. Table \ref{table:systematics} contains the full list of assumptions for each calibration.}
    \label{fig:FullPlots}
\end{figure}

The best fit functions from the near-term standard precision and future high precision calibration simulations are compared to the Monte Carlo true functions in Figure~\ref{fig:FullPlots}. Overall, the calibration shows the ability to fit the lower half of the efficiency function well, with a relatively larger uncertainty on the placement of the 100\% efficiency node. Calibration source rates are inherently less sensitive to the 80\% and 100\% efficiency nodes since fewer sources constrain the higher energies, and 80--100\% efficiency represents a smaller proportional increase than the lower efficiency nodes.

%\begin{figure}
    %\centering
    %\includegraphics[width=\textwidth]{TN_MCMC.png}
   % \caption{Plots for the best fit functions for thermal neutron simulated calibrations compared to the true efficiency function.  The blue lines represent the true efficiency functions of 400 eV with Gaussian width of 50 eV. Left column: standard precision calibration results with the set of systematic assumptions associated with near-term calibration for a dark matter search. 5\% uncertainty in source activity, 20\% uncertainty in thermal neutron simulations and 20\% uncertainty in photoneutron simulations and 0.49 background events/hour. A normalized NR spectrum from 2 GeV/$c^{2}$ dark matter in argon is shown in black. Right column: high precision simulated calibration results which assumes the systematics that could be attained by a calibration campaign to set up a future reactor CE$\nu$NS experiment. 1\% uncertainty in source strength, 5\% uncertainty in thermal neutron scattering rate, and 5\% uncertainty in photoneutron scattering rate and 0.49 background events/hour. These plots have a normalized reactor CE$\nu$NS spectra for comparison in green.}
  %  \label{fig:TN_curve}
%\end{figure}

Table \ref{table:CEvNSdata} shows the effective threshold and the uncertainty in source rate for each simulated calibration. Most of standard precision best fit functions for all four thresholds have effective thresholds within 20\% of the true threshold and are within 10\% in the expected dark matter rate. The high precision best fit functions are within the 5\% target for each threshold. Achieving a more aggressive 2\% or lower uncertainty goal on threshold, established as a best case by Ref.~\cite{SBCcevns}, would require further improvements to systematic uncertainties. The improvement from standard precision to high precision is primarily due to improved systematic uncertainties, with lower background having a smaller impact.

%For the 80 eV threshold, wthe 68\% confidence interval falls just outside the 20\% target, but the uncertainty is greatest for higher energy NR where the WIMP rate is lower, and the 80 eV threshold has the highest overall WIMP rate. Consequently, the uncertainty in WIMP rate is within the target. 

\begin{table}
\caption{The effective threshold and difference in signal rate for eight calibrations: standard precision calibrations and high precision calibrations at four different effective thresholds. The 80-320 eV thresholds calibrations use Thomson and photoneutron sources, while the 400 eV threshold uses neutron capture and photoneutron sources. The model effective threshold is calculated for best fit efficiency curve from each mock calibration, and the median of those values is presented. The dark matter or CE$\nu$NS rate difference is the percent disagreement between the signal bubble rate predicted by the Monte Carlo true efficiency function and the rate predicted by the best-fit model efficiency function, where a positive rate indicates a larger true signal. 2 GeV/$c^{2}$ mass dark matter is assumed. Uncertainties come from a symmetrized 68\% confidence interval over the mock datasets. This would contribute to the systematic error of a dark matter search or CE$\nu$NS experiment.}
\begin{center}
\begin{tabular}{  l l l  }
\hline
 & Standard Precision Calibration&\\
\hline
 True Threshold&  Model Effective Threshold& Dark Matter Rate Difference\\
\hline
\hline
  80 eV&  91 $\pm$ 17 eV&3.2 $\pm$ 5.9 \%\\ 
 160 eV&  169 $\pm$ 14 eV&2.1 $\pm$ 5.5 \%\\ 
 320 eV&  324 $\pm$ 63 eV&0.0 $\pm$ 10.0 \%\\
 400 eV&  418 $\pm$ 60 eV&3.7 $\pm$ 16.6 \%\\
 \hline
 & High Precision Calibration&\\
\hline
 True Threshold&  Model Effective Threshold& CE$\nu$NS Rate Difference\\
\hline
\hline
  80 eV&  80 $\pm$ 2 eV&-0.1 $\pm$ 0.6 \%\\ 
 160 eV&  161 $\pm$ 5 eV&-0.6 $\pm$ 1.9 \%\\ 
 320 eV&  321 $\pm$ 14 eV&-0.8 $\pm$ 3.8 \%\\
 400 eV&   405$\pm$ 37 eV&-0.9 $\pm$ 5.2 \%\\
 \hline
 \bottomrule
\end{tabular}
\label{table:CEvNSdata}
\end{center}
\end{table}
\section{Conclusion}
The SBC 10 kg liquid argon detector has the potential for world leading sensitivity to $\mathcal{O}$(GeV/$c^{2}$) mass dark matter and to perform a high statistics reactor CE$\nu$NS measurement. To calibrate this detector, the SBC Collaboration will combine a photoneutron scattering calibration with novel lower energy techniques including nuclear Thomson scattering and thermal neutron capture. These calibrations take advantage of noble liquid bubble chambers' unique set of features including electronic recoil rejection, low nuclear recoil threshold, and scintillation tagging of events. The scintillation tag gives neutron capture a unique appeal as a calibration technique. The calibration will also serve to validate simulations for nuclear Thomson scattering, which is expected to be a significant source of backgrounds for future low energy nuclear recoil rare event searches, including the SBC-SNOLAB experiment. 

The calibration campaign is simulated with mock datasets at four thresholds from 80 to 400~eV. A near term calibration is simulated with systematic uncertainties on simulation accuracy and source strength comparable to other recent bubble chamber calibrations. This calibration achieves the goal of 20\% or lower uncertainty in threshold. A higher precision calibration is also simulated, assuming dedicated effort to reduce sources of error, including using scintillation light collection to constrain source geometry, follow up calibrations, and shielding to reduce background if required. This calibration could reduce the uncertainty in threshold to 5\% or lower, as required to realize SBC physics reach in a reactor CE$\nu$NS measurement.

\acknowledgments
We gratefully acknowledge Rafi Sazzad for useful discussions regarding AmLi sources.

This work was supported by DOE Office of Science Grants No.\ DE-SC0015910, No.\ E-SC0017815, No.\ DE-SC0024254, and No.\ DE-SC0011702, National Science Foundation Grants No.\ DMR-1936432, No.\ PHY-2310112, and No.\ PHY-2411655, Project No.\ SECIHTI CBF-2025-I-1589, DGAPA UNAM Grants No.\ PAPIIT IN105923 and IN102326. We acknowledge support from the Canada First Research Excellence Fund through the Arthur B. McDonald Canadian Astroparticle Physics Research Institute and the Natural Sciences and Engineering Research Council of Canada (NSERC). 

We also thank the Digital Research Alliance of Canada \cite{computecanada} and the Centre for Advanced Computing, ACENET, Calcul Qu\'ebec, Compute Ontario, and WestGrid for computational support. The work of D.~Durnford and A.~de~St.~Croix was supported by the NSERC Canada Graduate Scholarships -- Doctoral program.

\clearpage
\appendix
\addcontentsline{toc}{section}{Appendices}
\textbf{\huge Appendices}

\section{JAEA GEANT4.10.5.1 Correction} \label{JAEA changes}

The corrections to the JAEA cross-section implementation involve two separate fixes to the file \path{G4JAEAElasticScatteringModel.cc}.

First, in converting the differential cross-section to a probability distribution function (pdf) in scattering angle $\theta$, a factor of $2\pi \sin{\theta}$ was missing:

\begin{equation}
\frac{d\sigma}{d\theta}=2\pi\sin{\theta}\frac{d\sigma}{d\Omega}.
\end{equation}

The second fix relates to the cross-section tabulation being in one-degree bins in $\theta$. This is good enough for the large angle scatters relevant for this work, since the cross-section varies slowly in this region, but the cross-section varies steeply within the first degree due to the dominance of the Rayleigh scattering process. As a result, the sum of the calculated pdf differs from the total cross-section by a large factor. To attempt to address this, the pdf was normalized to agree with the total cross-section, but that results in substantial errors in the large angle scattering probability. The code was modified to instead adjust only the one-degree scattering probability as necessary to bring the sum of the pdf and the total cross-section into agreement, retaining accurate probabilities for large angle scattering.\footnote{The original code and the updated code for the G4JAEAElasticScatteringModel.cc are available in Github. The original can be found \url{https://github.com/Geant4/geant4/blob/geant4-10.5-release/source/processes/electromagnetic/lowenergy/src/G4JAEAElasticScatteringModel.cc} and a short repository with the updated file and screenshot comparisons to the changed code can be found \url{https://github.com/NoahRLamb/GEANT4.10.5.1_JAEA_correction}} Simulations using the revised code have been validated to within a few percent against the calculated cross-section calculations.

\section{Derivation of Thermal Neutron Capture Spectrum}
\subsection{Stopping of keV Argon Recoils} \label{Stopping of keV Argon Recoils}
In Equation \ref{eq:TN_E_r}, the amount of energy released between the emission of $\gamma_i$ and $\gamma_{i+1}$ is referred as $\Delta E_i$, which can in turn be written as:
\begin{equation}
\label{eq:TN_del_Ei}
    \Delta E_{i}=E_{ki}-E_k(E_{ki},\Delta t_i).
\end{equation}
Here, $E_{ki}$ is the kinetic energy after the $i$-th gamma emission and $E_k(E_{ini},t)$ is a function that describes the remaining kinetic energy of the nucleus after some time $t$ starting with an initial energy $E_{ini}$.

To determine for the function $E_k(E_{ini},t)$, the Lindhard, Scharfff, and Schott (LSS) theory is utilized to describe stopping power in matter. LSS splits $dE/dx$ into the sum of nuclear and electronic stopping \cite{Mangiarotti2007, Ziegler2010} :
\begin{equation}
\label{eq:stoppingpower}
    \frac{dE}{dx} = \frac{dE_\mathrm{N}}{dx} + \frac{dE_e}{dx}.
\end{equation}
The nuclear stopping is given by: 
\begin{equation}
\label{eq:LLS1}
   \frac{\mathrm{d} E_\mathrm{N}}{\mathrm{d}x}=NS_n(E_\mathrm{N}),
\end{equation}
\begin{equation}
\label{eq:LLS2}
   S_n(E_\mathrm{N}) =\pi Z_1 Z_2 e^2 M_1\frac{ln(1+a \epsilon)}{2(\epsilon+b\epsilon^c+d\epsilon^{0.5})},
\end{equation}
\begin{equation}
\label{eq:LLS3}
   \epsilon\coloneqq \frac{a_U M_2 E_\mathrm{N}}{Z_1 Z_2 e^2 (M_1+M_2)}.
\end{equation}

Here, $N$ is the liquid nuclear number density, $S_n$ is the nuclear dissipation rate per nucleus, $\epsilon$ is the reduced energy, $a_U = 0.8853 \, a_0 \, /({Z_1^{0.23} + Z_2^{0.23}})$ is the universal screening length and $a_0$ is Bohr radius. $M_1$ is the atomic mass of the projectile ion, $M_2$ is the atomic mass of the target ion, $Z_1$ is the atomic number of the projectile ion, and $Z_2$ is atomic number of the target atom. For liquid argon, the constants are $a=1.1383$, $b=0.01321$, $c=0.21226$, and $d=0.19593$. 

Regarding $dE_e/dx$, the Stopping and Range of Ions in Matter (SRIM) package \cite{Ziegler2010} provides tabulated values of the electronic stopping power as a function of kinetic energy. At low energies, these values are proportional to the velocity, or the square root of $E$:  
\begin{equation}
    \label{eq:ED1}
   \frac{\mathrm{d} E_e}{\mathrm{d}x}=k\sqrt{E_e},
\end{equation}
where k = 2.18$\times10^{-2}\pm 1.81 \times 10^{-6} \sqrt{\rm{eV}}$/\text{\AA}, obtained by a fit to the tabulated data.

\begin{figure}
\centering
\includegraphics [width=0.6\textwidth] {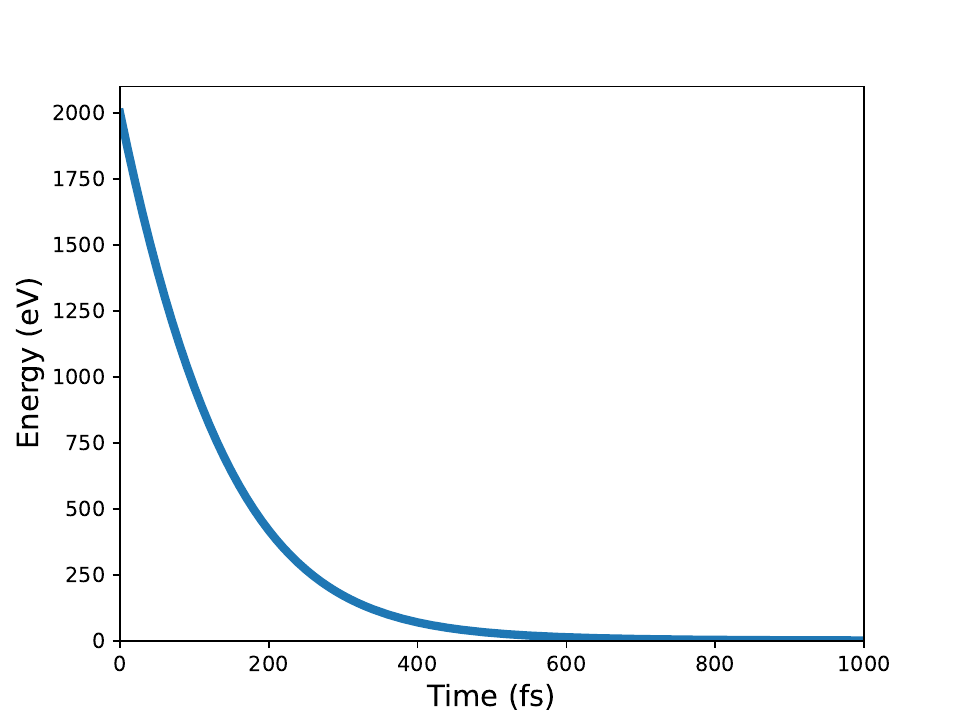} 
\caption{Nuclear recoil energy in liquid argon as a function of time since collision, taking an initial energy $E_{ini}=2$ keV.}
\label{fig:TNdEdt}
\end{figure}

With $dE/dx$ obtained as a function of the energy, $E$, $dE/dt$ is found by:
\begin{equation}
    \delta E = \frac{dE}{dx} \delta x = \frac{dE}{dx} \frac{dx}{dt} \delta t = \frac{dE}{dx} \sqrt{\frac{2E}{M}} \delta t, 
    \end{equation}
    \begin{equation}
         \frac{dE}{dt} = \frac{dE}{dx} \sqrt{\frac{2E}{M}}.
\end{equation}
This differential equation can be solved numerically to obtain $E_k(E_{ini},t)$.  Figure  \ref{fig:TNdEdt} shows $E_k$ as a function of time for $E_{ini}$ = 2 keV.

\subsection{De-excitation Configuration and MC simulations} \label{De-excitation Configuration and MC simulations}
Next, the gamma energies $\gamma_i$ and lifetimes $\Delta t_{i}$ for $^{41}$Ar and $^{37}$Ar are obtained from the ENSDF \cite{ENSDF1, ENSDF2} database. The energy imparted by the $i$-th gamma decay, $E_{ki}$, is calculated by conservation of momentum: 
\begin{equation}
\label{eq:MC_Momentum}
   \vec{p}'_{\mathrm{Ar},i} = \vec{p}_{\mathrm{Ar}, i} +\vec{p}_{\gamma, i},
\end{equation}
where $\vec{p}'_{\mathrm{Ar},i}$ is the argon nucleus momentum before the emission of $\gamma_i$, $\vec{p}_{\mathrm{Ar},i}$ is the argon nucleus momentum after emission of $\gamma_i$, and $\vec{p}_{\gamma, i}$ is the momentum of $\gamma_i$ assuming isotropic gamma-ray emission.
Combining the above considerations, 
%the function $E_k(E_{ini},t)$ is displayed in the Figure \ref{fig:TNdEdt}
the final argon recoil spectrum after thermal neutron capture is obtained and displayed in Figure \ref{fig:TN_spectrum}.

\subsection{Uncertainty Evaluation} \label{Uncertainty Evaluation}
\begin{figure}
\centering
\begin{subfigure}{.54\textwidth}
  \centering
  \includegraphics[width=\linewidth]{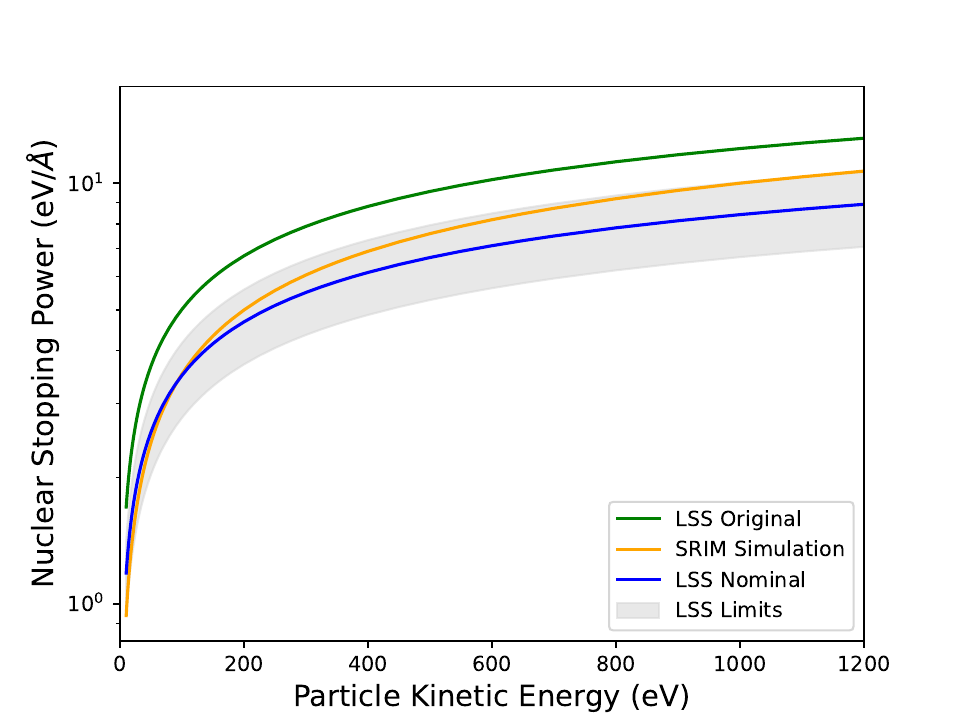}

  \caption{}
  \label{fig:SRIM_limits}
\end{subfigure}%
\begin{subfigure}{.54\textwidth}
  \centering
  \includegraphics[width=\linewidth]{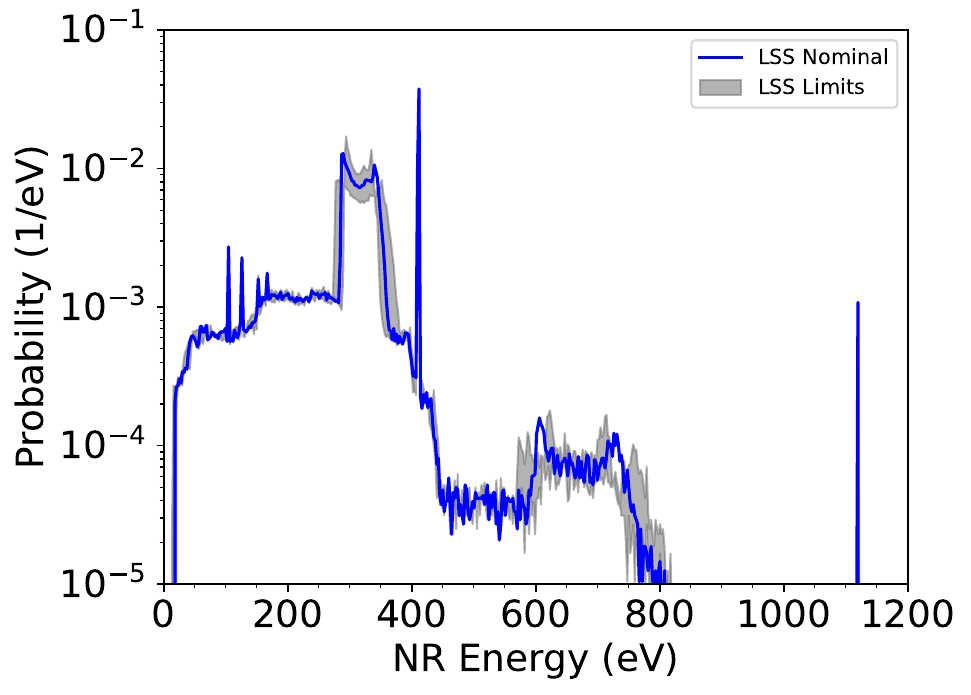}
  \caption{ }
  \label{fig:spectrum_limits_LSS}
\end{subfigure}%

\begin{subfigure}{.53\textwidth}
  \centering
  \includegraphics[width=\linewidth]{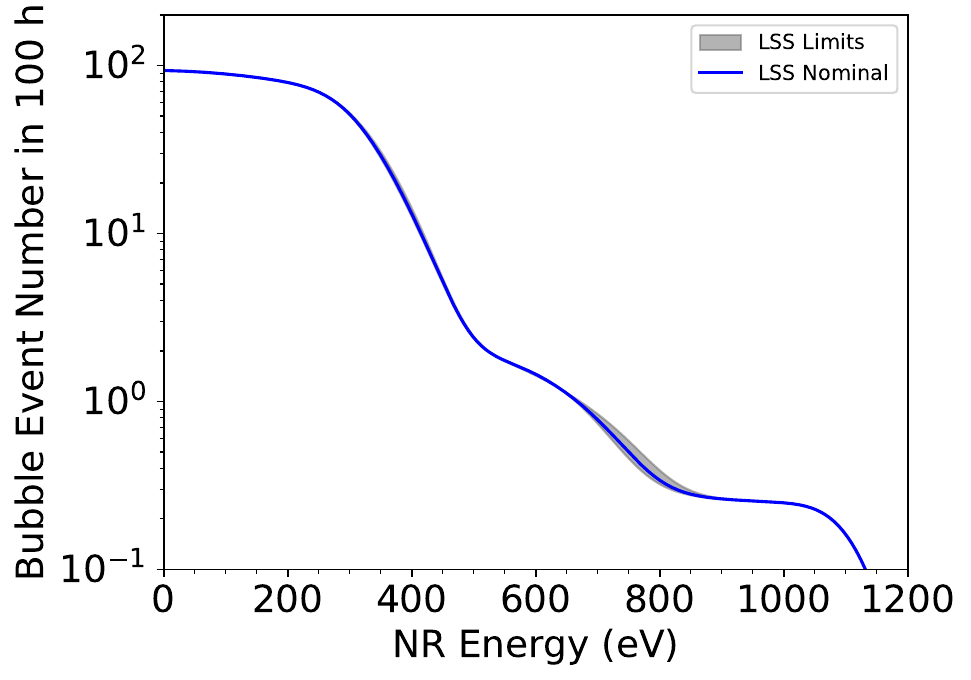}
  \caption{}
  \label{fig:Bubble_num_LSS}
\end{subfigure}
\caption{Uncertainty estimation from the LSS nuclear stopping model. (a) Nuclear stopping power comparison between SRIM data and the LSS model. The original LSS curve shows the unscaled analytical prediction and the orange curve shows the SRIM prediction. The ``LSS nominal'' curve shows the LSS analytic prediction scaled to match SRIM in the energy range of interest. The shaded gray band represents the range obtained by applying upper and lower scaling factors to the LSS model within the 0–2 keV recoil energy range.    (b) Variation in the predicted argon energy spectrum from varying the LSS scaling factor. (c) The predicted  bubble rate from thermal neutron capture as a function of the NR energy threshold, assuming the nucleation efficiency curve is a NCDF function with a width of 50 eV. The gray band shows the uncertainty on the predicted rate caused by varying the LSS scaling. }
\label{fig:SRIM_uncertantiy1}
\end{figure}

The SRIM calculations of total ion stopping power are shown by the orange line in Fig.~\ref{fig:SRIM_limits}. As described above, Eqs. \ref{eq:stoppingpower}, \ref{eq:LLS1}, and \ref{eq:ED1} from the LSS model are used to derive an easier-to-use analytic expression for the stopping power, represented by the blue line in Fig.~\ref{fig:SRIM_limits}. A linear scale factor is applied to the LSS calculation to better match the SRIM simulation in the primary energy range of interest of 0--1200 eV, giving the green line; this is the stopping power that is used in the final calculation to obtain Fig.~\ref{fig:TN_spectrum}. 

To estimate a systematic uncertainty in the calculation, the scale factor is varied, shown by the gray uncertainty band in Figure \ref{fig:SRIM_uncertantiy1}. Running the calculation forward, the range of output spectra obtained at the extreme limits of the gray band is shown in Figure \ref{fig:spectrum_limits_LSS}. In particularly, a widening of the spectrum around the peaks at $\sim$300 eV and $\sim$600--800 eV can be seen. 

The variation in the predicted bubble rate as a function of the NR energy threshold is shown in Figure \ref{fig:SRIM_limits}, under the assumption that the nucleation efficiency curve is a NCDF function with 50 eV width(as in the main analysis). The result shows minimal effects of this systematic. The final uncertainty on the reconstructed value of the 50\% efficiency node energy caused by LSS scaling is less than 1\% at 400 eV threshold.  

\begin{figure}
\centering
\begin{subfigure}{.5\textwidth}
  \centering
  \includegraphics[width=1.\linewidth]{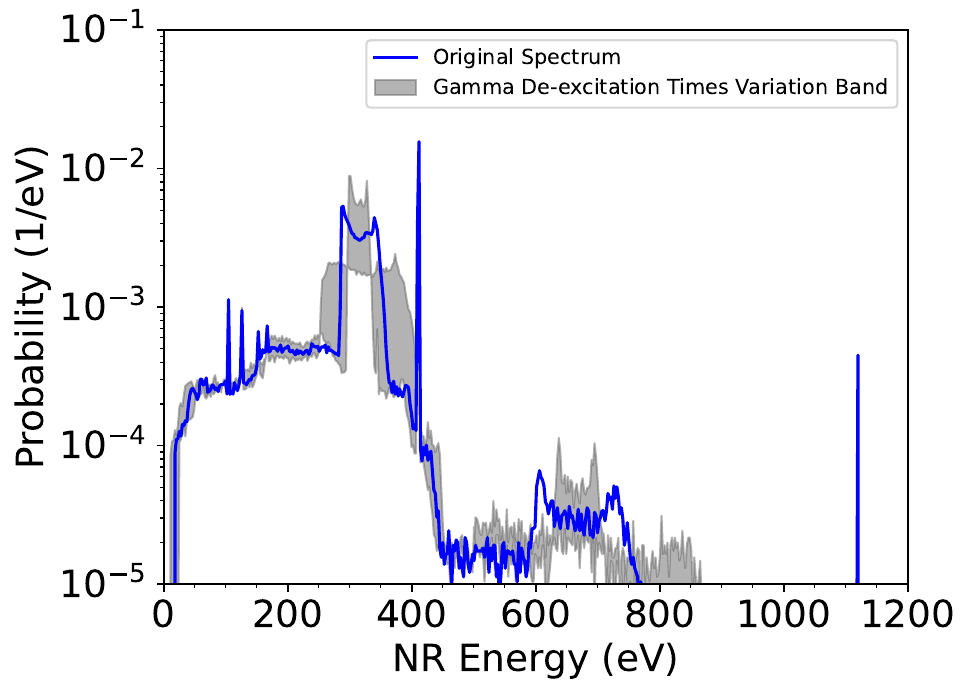}
  \caption{ }
\end{subfigure}%
\begin{subfigure}{.5\textwidth}
  \centering
  \includegraphics[width=1.\linewidth]{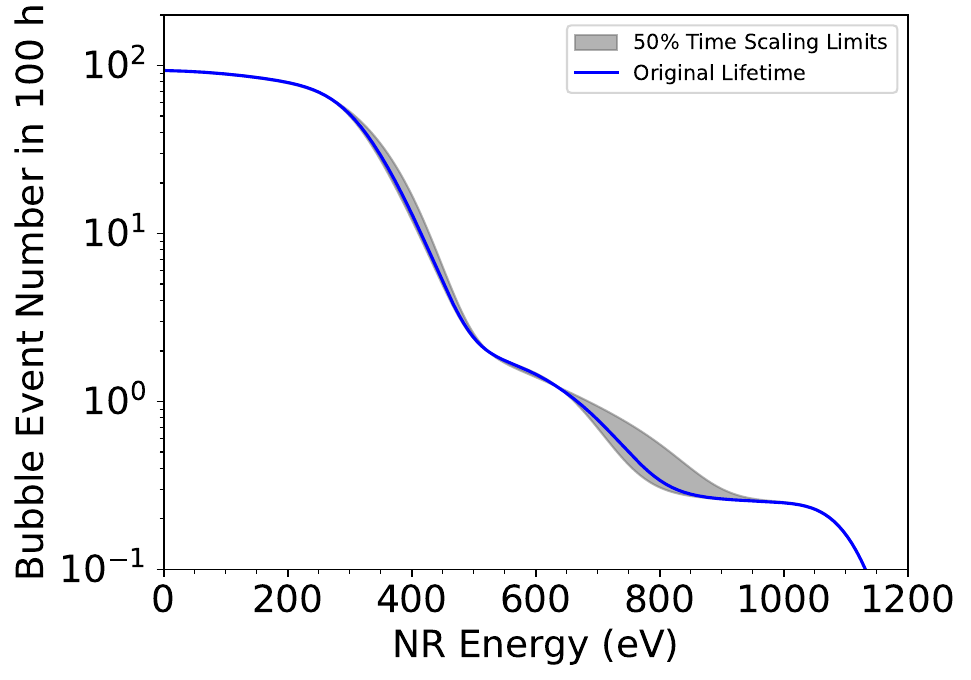}
  \caption{}
\end{subfigure}
\caption{Simulation uncertainty estimation from gamma de-excitation time. (a) Variation in the predicted argon energy spectrum from varying the main gamma ray lifetimes by 50\%.  (b) The predicted bubble rate from thermal neutron capture as a function of the NR energy threshold, assuming the nucleation efficiency curve is a NCDF function with a width of 50 eV. The gray band shows the uncertainty on the predicted rate caused by varying the gamma ray decay times.}
\label{fig:SRIM_uncertantiy2}
\end{figure}

The effect of uncertainty on the gamma-ray lifetime tabulated by ENSDF is also checked. These lifetimes are varied by a factor of 50\% with a similar impact on the calculated recoil spectrum, as shown in Figure \ref{fig:SRIM_uncertantiy2}. The estimated systematic uncertainty on the reconstructed value of the 50\% efficiency node due to imperfect knowledge of the gamma ray decay times is less than 4\% of at 400 eV threshold.

\clearpage

\clearpage
\bibliography{main.bib}
\end{document}